\documentclass[aps,prx,10pt,showpacs,eqsecnum,amsmath,twocolumn,superscriptaddress,tightenlines,floatfix,notitlepage,longbibliography]{revtex4-1}
\usepackage[dvipsnames]{xcolor}
\usepackage{amsmath,amssymb}
\usepackage{amsfonts}
\usepackage{graphicx}
\usepackage{mathtools}
\allowdisplaybreaks[1]
\usepackage[normalem]{ulem}
\usepackage{soul}
\usepackage[colorlinks=true,linkcolor=NavyBlue,urlcolor=NavyBlue,citecolor=NavyBlue,breaklinks]{hyperref}
\newcommand{\lapprox}{%
\mathrel{%
\setbox0=\hbox{$<$}
\raise0.6ex\copy0\kern-\wd0
\lower0.65ex\hbox{$\sim$}
}}
\usepackage[normalem]{ulem}

\newcommand{\intall}{\int_{-\infty}^{\infty}}
\newcommand{\ket}[1]{|#1\rangle}

\newcommand{\avg}[1]{\langle#1\rangle}
\newcommand{\Avg}[1]{\left\langle#1\right\rangle}

\newcommand{\sinc}{\operatorname{sinc}}

\newcommand{\erfc}{\operatorname{erfc}}

\newcommand{\bs}[1]{\boldsymbol{#1}}
\newcommand{\abs}[1]{\left|#1\right|}
\newcommand{\bk}[1]{\left(#1\right)}
\newcommand{\Bk}[1]{\left[#1\right]}
\newcommand{\BK}[1]{\left\{#1\right\}}

\newcommand{\trace}{\operatorname{tr}}

\newcommand{\nn}{\nonumber \\}
\newcommand{\expect}{\mathbb E}
\newcommand{\dig}{\psi}

\begin{document}
\title
{The quantum Bell-Ziv-Zakai bounds and Heisenberg limits for waveform estimation}

\author{Dominic W. Berry}
\affiliation{Department of Physics and Astronomy, Macquarie University, NSW 2109, Australia}

\author{Mankei Tsang}
\affiliation{Department of Electrical and Computer Engineering,
  National University of Singapore, 4 Engineering Drive 3, Singapore
  117583}
\affiliation{Department of Physics, National University of Singapore,
  2 Science Drive 3, Singapore 117551}

\author{Michael J. W. Hall}
\author{Howard M. Wiseman}
\affiliation{Centre for Quantum Computation and Communication Technology (Australian Research Council), Centre for Quantum Dynamics, Griffith University, Brisbane, QLD 4111, Australia}

\date{\today}

\begin{abstract}
  We propose quantum versions of the Bell-Ziv-Zakai
  lower bounds on the error in multiparameter
  estimation.  As an application we consider measurement of a
  time-varying optical phase signal with stationary Gaussian prior
  statistics and a power law spectrum $\sim 1/|\omega|^p$, with
  $p>1$. With no other assumptions, we show that the mean-square error
  has a lower bound scaling as $1/{\cal N}^{2(p-1)/(p+1)}$, where
  ${\cal N}$ is the time-averaged mean photon flux. Moreover, we show
  that this accuracy is achievable by sampling and interpolation, for
  any $p>1$. This bound is thus a rigorous generalization of the
  Heisenberg limit, for measurement of a single unknown optical phase,
  to a stochastically varying optical phase.
\end{abstract}

\maketitle
\section{Introduction}
The probabilistic nature of quantum mechanics imposes fundamental
limits to hypothesis testing and parameter estimation
\cite{helstrom,glm_science,glm2011,holevo12}. Such limits are relevant
to many metrological applications, such as optical interferometry,
optomechanical sensing, gravitational-wave detection
\cite{braginsky,twc,tsang_nair,tsang_open}, optical imaging
\cite{treps,centroid,taylor2013}, magnetometry, gyroscopy, and atomic
clocks \cite{bollinger}. The ultimate quantum limits to parameter estimation have
been studied extensively in recent years, as they imply
that a minimum amount of resource, such as the average photon number
for optical phase estimation, is needed to achieve a desired
precision, regardless of the measurement method.

For the measurement of a single optical phase parameter, the ultimate quantum limit 
to the mean-square error scales as $1/\bar{n}^2$, where $\bar{n}$
is the average photon number of the field which undergoes that phase shift. 
This scaling is often called the Heisenberg
limit. After years of speculation and debate
\cite{yurke,sanders,ou,bollinger,zwierz,zwierz_err,rivas,luis_rodil,luis13,anisimov,zhang13},
the Heisenberg limit for single-parameter linear phase estimation has 
only recently been proven
\cite{qzzb,glm2012,berry12,hall2012,nair2012,gm_useless,hall_prx}. Although
decoherence, such as optical loss and dephasing, can impose stricter
limitations
\cite{knysh,escher,escher_bjp,escher_prl,latune,demkowicz,tsang_open,knysh14},
the Heisenberg limit is a more fundamental bound and will be
increasingly relevant as quantum technologies continue to improve and
decoherence effects are further reduced.

Many real-world tasks, such as optical imaging \cite{kolobov,humphreys},
quantum tomography and system identification
\cite{paris_rehacek,young}, and waveform estimation (e.g.\ estimating a 
signal that varies continuously in time)
 \cite{twc,tsang_open,bhw2013,wheatley,yonezawa,iwasawa},
require the estimation of multiple parameters.  Multiparameter quantum
Cram\'er-Rao bounds have been known since the 1970s 
\cite{yuen_lax,helstrom_kennedy,paris,twc}, but efforts to derive
multiparameter Heisenberg limits from these bounds have not been successful.
This is not surprising, since even in the case of single-parameter phase estimation, 
it is not possible, without additional assumptions on the state, 
to derive the Heisenberg limit from the quantum Cram\'er-Rao 
bound. (The latter gives a lower bound on the mean-square error of  
$1/(\Delta n)^2$, which does not imply the Heisenberg limit of $1/\bar{n}^2$,
as can be seen from the state $\frac{\sqrt{3}}{2}\sum_{n=0}^\infty 2^{-n}\ket{2^n}$
which has $\bar{n}=3/2$ but divergent $\Delta n$.)
	
In Ref.~\cite{bhw2013}, some of us
recently proposed a Heisenberg-style limit for the estimation of an
optical phase waveform with stationary Gaussian prior statistics and a
power-law spectrum. However, that limit, being derived from a quantum
Cram\'er-Rao bound, requires additional assumptions: it applies 
only to the specific class of optical beams described by Gaussian fields, with statistics that are both 
stationary and time-symmetric.
A very different approach was that of Ref.~\cite{zhang14},
which derives a multiparameter Heisenberg limit for independent
parameters by applying the single-parameter Heisenberg limit to each
parameter.  In practice, multiple parameters often have nontrivial prior correlations, 
particularly in the case of continuous waveform estimation,
where the correlations are crucial to pose the problem \cite{vantrees}. Thus the existence of general 
Heisenberg limits for such cases has remained an open question.

In this paper, we derive new quantum bounds on multiparameter
estimation by developing quantum versions of the
Bell-Ziv-Zakai bounds \footnote{The terminology ``Bell-Ziv-Zakai
  bounds'' was adopted in \cite{bell}.}.  The Bell-Ziv-Zakai bounds
were proposed in 1997 by Bell \textit{et al.}\ \cite{bell1997},
building upon the Ziv-Zakai bound \cite{ziv-zakai}, and futher
generalized by Basu and Bresler \cite{basu2000}.  We then apply our
bounds to the notable task of quantum optical phase waveform
estimation.  Here, the waveform to be estimated is a time-varying
phase shift signal, $X(t)$, applied to an optical beam.  For a
waveform $X(t)$ with stationary Gaussian prior statistics and a
power-law spectrum ($\propto 1/|\omega|^p, p > 1$), we prove a lower
bound on the mean-square error with a $1/{\cal N}^{2(p-1)/(p+1)}$
scaling, where ${\cal N}$ is the mean photon flux. This proof confirms
that the scaling previously proposed in Ref.~\cite{bhw2013} is valid
for arbitrary quantum states. Moreover, we show that this scaling is
achievable for all $p>1$. Previously, achievability has been shown
only numerically, and only for $p=2$~\cite{bw2013}.  By contrast the
results in the current paper are completely rigorous Heisenberg
bounds, being both applicable to arbitrary field states, and
achievable, for all $p>1$.

This paper is separated into two main parts. The first part,
Sec.~\ref{QZZB} to Sec.~\ref{sec:ach}, assumes unbounded parameters
and focuses on the mean-square error as the distortion measure
(i.e.\ the figure of merit for the accuracy of the estimation).
The second part, Sec.~\ref{sec:per} and Sec.~\ref{sec:ach2}, focuses
on periodic distortion functions, which are more appropriate for
periodic parameters such as phase or orientation angles for
gyroscopy. They are also insensitive to
phase-wrap errors and enable us to rigorously prove
that our bounds are achievable.

\section{Quantum Bell-Ziv-Zakai bounds}
\label{QZZB}
\subsection{Classical estimation}
First we summarize known results for the classical estimation problem, then present quantum versions of the bounds in Sec.~\ref{sec:qbzzb}.
Let $\bs X$ be a column vector of unknown real parameters, $P_{\bs X}(\bs
x)$ be the prior probability density, and $P_{\bs Y|\bs X}(\bs y|\bs x)$
be the likelihood function with observation $\bs Y$.
Both $\bs X$ and $\bs Y$ are random variables.
Note that $\bs Y$ need
not be the same dimension as $\bs X$. Further, let $\check{\bs X}(\bs Y)$ be the estimator of $\bs X$ from $\bs Y$. 
(We use $\check{\bs X}$ rather than $\hat{\bs X}$, as is common in statistics, to avoid possible confusion 
with quantum operators.) We also define the
error vector as
\begin{equation}
\bs \epsilon(\bs X,\bs Y) := \check{\bs X} (\bs Y) - \bs X .
\end{equation}
To characterise the performance of the estimate, we consider
a distortion function of the form $D(\bs u^\top \bs
\epsilon)$, where $\bs u$ is a given but arbitrary real column vector that
defines the error components of interest, and $\top$ denotes the
transpose. For example, the mean-square error for a particular
component $X_k$ is the expected value of a distortion function
$D(x)=x^2$ with $u_j = \delta_{jk}$, so
\begin{equation}
D(\bs u^\top\bs \epsilon) = \Bk{\check X_k(\bs Y)-X_k}^2.
\label{point}
\end{equation}
Suppose that the distortion function is symmetric [that is, 
$D(\bs u^\top\bs \epsilon)=D(|\bs u^\top\bs \epsilon|)$],
nondecreasing on $[0,\infty)$,
differentiable, and has $D(0) = 0$. Then the expected distortion is,
from Eq.~(44) of Ref.~\cite{bell1997},
\begin{align}
\expect\Bk{D(|\bs u^\top \bs \epsilon|)} &= \frac{1}{2}
\int_0^\infty d\tau \, \dot D\bk{\frac{\tau}{2}}
\textrm{Pr}\bk{|\bs u^\top\bs\epsilon| \ge \frac{\tau}{2}},
\label{D_bound}
\end{align}
where $\dot{D}$ is the derivative of $D$, $\expect$  denotes 
expectation over $\bs X$ and $\bs Y$, and
Pr is the probability for the Boolean function of these random variables to be true.
For a general 
mean-square error criterion, the expected distortion can be expressed in terms of
the error covariance matrix $\bs\Sigma$ as
\begin{align}
\expect\Bk{D(|\bs u^\top \bs \epsilon|)}
&=   \expect\Bk{(\bs u^\top\bs\epsilon)^2}   = \bs u^\top\bs\Sigma\bs u,
\nn
\bs\Sigma &:=\expect\bk{\bs\epsilon\bs\epsilon^\top}.
\label{quad_D}
\end{align}
Since $\dot D$ is assumed to be nonnegative, a lower bound on the
expected distortion can be obtained by lower-bounding the probability
$\textrm{Pr}\bk{|\bs u^\top\bs\epsilon| \ge \tau/2}$.
  Using Eqs.~(31) and (35) of \cite{bell1997} to bound
  $\textrm{Pr}\bk{|\bs u^\top\bs\epsilon| \ge \tau/2}$ and noting
  Property 1 of \cite{bell1997}, yields the
  Bell-Ziv-Zakai bounds \cite{bell,bell1997}:
\begin{align}
&\expect\Bk{D(|\bs u^\top \bs \epsilon|)} 
\ge 
\frac{1}{2}
\int_0^\infty d\tau \, \dot D\bk{\frac{\tau}{2}}
\mathcal V\bigg\{
\max_{\bs v:\bs u^\top\bs v = 1} \int d\bs x
\nn&\quad\times
\Bk{P_{\bs X}(\bs x)+P_{\bs X}(\bs x+\bs v\tau)}
P_e(\bs x,\bs x+\bs v\tau)\bigg\},
\label{bzzb1}
\\ &\ge
\int_0^\infty d\tau \, \dot D\bk{\frac{\tau}{2}}
\mathcal V\bigg\{\max_{\bs v: \bs u^\top \bs v = 1} \int d\bs x
\nn&\quad\times
\min\Bk{P_{\bs X}(\bs x),P_{\bs X}(\bs x + \bs v\tau)}
P_e^{\rm el}(\bs x,\bs x+\bs v\tau)\bigg\} .
\label{bzzb2}
\end{align}
Here $\mathcal V$ is the valley-filling function defined as
\begin{align}
\mathcal V \{f(\tau)\} := \max_{\eta: \eta \ge 0} f(\tau+\eta),
\end{align}
and $P_e(\bs x,\bs x+\bs v\tau)$ is the minimum error probability for
the Bayesian binary hypothesis testing problem with hypotheses defined
as $\mathcal H_0$ and $\mathcal H_1$, observation probability
densities given by $P(\bs y|\mathcal H_0) = P_{\bs Y|\bs X}(\bs y|\bs
x)$ and $P(\bs y|\mathcal H_1) = P_{\bs Y|\bs X}(\bs y|\bs x+\bs
v\tau)$, and prior probabilities given by
\begin{align}
\pi_0 &:= \operatorname{Pr}(\mathcal H_0) = 
\frac{P_{\bs X}(\bs x)}{P_{\bs X}(\bs x)+P_{\bs X}(\bs x+\bs v\tau)},
\label{P0}
\\
\pi_1 &:= \operatorname{Pr}(\mathcal H_1) = 1-\operatorname{Pr}(\mathcal H_0).
\label{P1}
\end{align}
To be explicit
\cite{kailath1967,toussaint},
\begin{align}
&P_e(\bs x,\bs x+\bs v\tau) 
\nn
&= 
\frac{1}{2}-\frac{1}{2}\int d\bs y
\left|\pi_0 P_{\bs Y|\bs X}(\bs y|\bs x)-
\pi_1 P_{\bs Y|\bs X}(\bs y|\bs x+\bs v\tau)\right|.
\label{Pe}
\end{align}
$P_e^{\rm el}$ is defined in the same way as $P_e$ except that the prior
probabilities are equal ($\pi_0=\pi_1 = 1/2$).  

\subsection{Quantum estimation}
\label{sec:qbzzb}
For the quantum parameter estimation problem, let $\rho_{\bs x}$ be
the density operator that describes
the state of a quantum probe as a function of the unknown parameter
$\bs x$, and $E(\bs y)$ be the positive operator-valued measure (POVM)
that describes the measurement with outcome $\bs y$
\cite{helstrom}. The likelihood function becomes
\begin{align}
P_{\bs Y|\bs X}(\bs y|\bs x) &= \trace [ E(\bs y)\rho_{\bs x} ],
\end{align}
with $\trace$ denoting the operator trace. It is known
\cite{helstrom,gm_useless} that, for any POVM,
\begin{align}
P_e(\bs x,\bs x+\bs v\tau)
&\ge \frac{1}{2} - \frac{1}{2}\big{|}\big{|}\pi_0\rho_{\bs x}-
\pi_1\rho_{\bs x+\bs v\tau}\big{|}\big{|}_1
\label{trace_bound}
\\
&\ge \frac{1}{2}\Bk{1 - \sqrt{1-4\pi_0\pi_1F(\bs x,\bs x+\bs v\tau)}},
\label{F_bound}
\end{align}
where 
\begin{align}
||A||_1 &:= \trace \sqrt{A^\dagger A}
\end{align}
is the trace norm and
\begin{align}
F(\bs x,\bs x+\bs v\tau) &:= 
\bk{\trace \sqrt{\sqrt{\rho_{\bs x}}\rho_{\bs x+\bs v\tau}\sqrt{\rho_{\bs x}}}}^2
\label{fidelity}
\end{align}
is the Uhlmann fidelity.  Equations~\eqref{trace_bound} and
\eqref{F_bound}, together with the first Bell-Ziv-Zakai bound given
by Eq.~(\ref{bzzb1}), then give quantum lower bounds on the estimation
error:
\begin{align}
&\expect\Bk{D(|\bs u^\top \bs \epsilon|)} 
\ge 
\frac{1}{4}
\int_0^\infty d\tau \, \dot D\bk{\frac{\tau}{2}}
\nn&\quad\times
\mathcal V\bigg\{
\max_{\bs v:\bs u^\top\bs v = 1} \int d\bs x\Bk{P_{\bs X}(\bs x)+P_{\bs X}(\bs x+\bs v\tau)}
\nn&\quad\times
\left(1 - \big{|}\big{|}\pi_0\rho_{\bs x}-
\pi_1\rho_{\bs x+\bs v\tau}\big{|}\big{|}_1\right)
\bigg\}
\label{qbzzb1}
\\
&\ge 
\frac{1}{4}
\int_0^\infty d\tau \, \dot D\bk{\frac{\tau}{2}}
\nn&\quad\times
\mathcal V\bigg\{
\max_{\bs v:\bs u^\top\bs v = 1} \int d\bs x\Bk{P_{\bs X}(\bs x)+P_{\bs X}(\bs x+\bs v\tau)}
\nn&\quad\times
\Bk{1 - \sqrt{1-4\pi_0\pi_1F(\bs x,\bs x+\bs v\tau)}}
\bigg\}.
\end{align}
Similarly, Eq.~(\ref{bzzb2}) and quantum bounds on $P_e^{\rm el}$ via
Eqs.~(\ref{trace_bound}) and (\ref{F_bound}) lead to the lower  bounds 
\begin{align}
&\expect\Bk{D(|\bs u^\top \bs \epsilon|)} 
\ge \frac{1}{2}
\int_0^\infty d\tau \, \dot D\bk{\frac{\tau}{2}}
\nn&\quad\times
\mathcal V\bigg\{\max_{\bs v: \bs u^\top \bs v = 1}
\int d\bs x \min\Bk{P_{\bs X}(\bs x),P_{\bs X}(\bs x + \bs v\tau)}
\nn&\quad\times
\left(1 - \frac{1}{2}\big{|}\big{|}\rho_{\bs x}-
\rho_{\bs x+\bs v\tau}\big{|}\big{|}_1\right)
\bigg\}
\\
&\ge \frac{1}{2}
\int_0^\infty d\tau \, \dot D\bk{\frac{\tau}{2}}
\nn&\quad\times
\mathcal V\bigg\{\max_{\bs v: \bs u^\top \bs v = 1}
\int d\bs x \min\Bk{P_{\bs X}(\bs x),P_{\bs X}(\bs x + \bs v\tau)}
\nn&\quad\times
\Bk{1 - \sqrt{1-F(\bs x,\bs x+\bs v\tau)}}
\bigg\}.
\label{qbzzb4}
\end{align}
We call Eqs.~(\ref{qbzzb1})--(\ref{qbzzb4}) the
quantum Bell-Ziv-Zakai bounds. 

\begin{figure}[htbp]
\centerline{\includegraphics[width=0.45\textwidth]{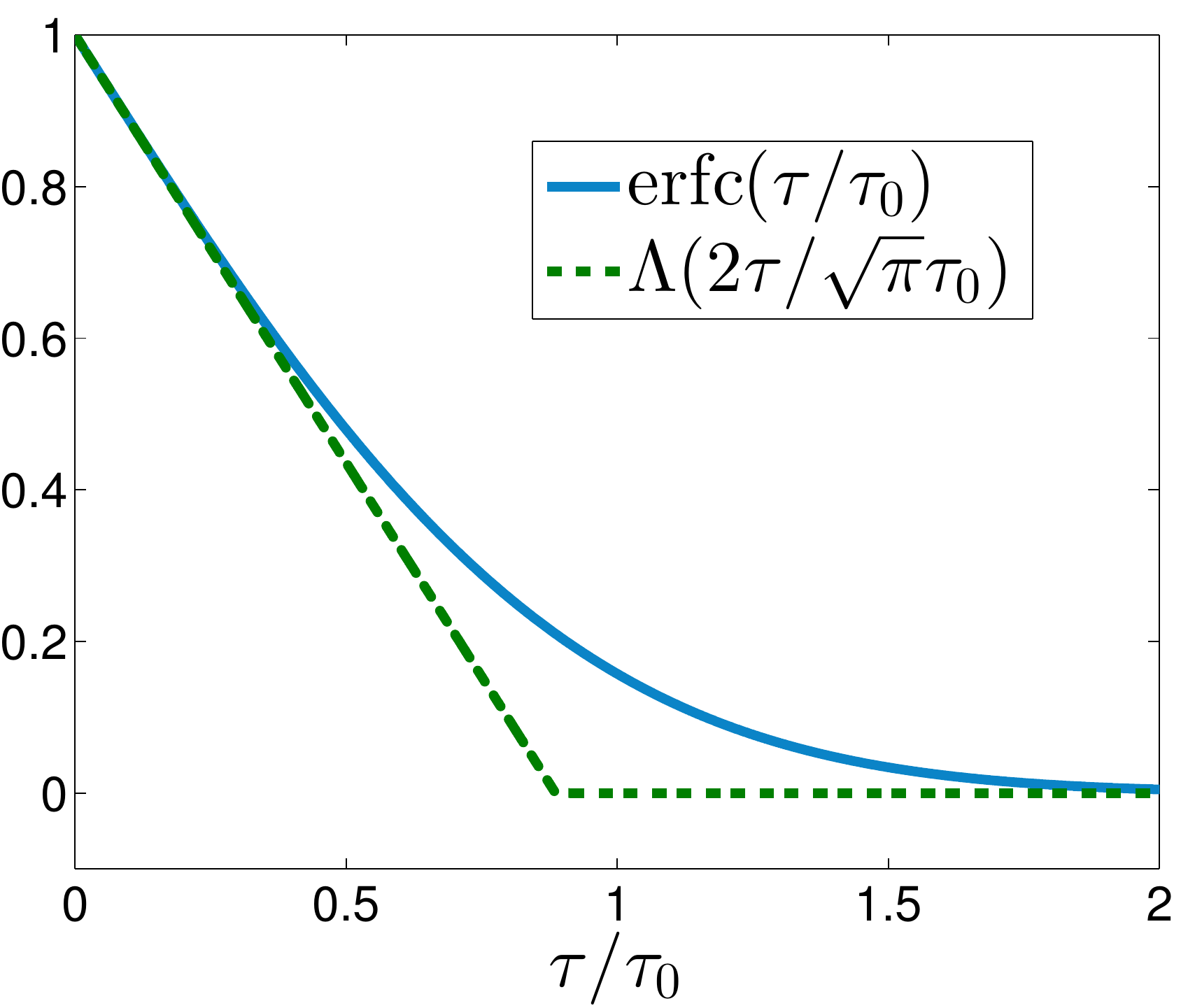}}
\caption{The erfc function and a lower bound using the triangle
  function $\Lambda$.}
\label{fig:erfc}
\end{figure}

To derive further analytic results, we focus on the fidelity bound
given by Eq.~(\ref{qbzzb4}). It can be further simplified if $F(\bs
x,\bs x+\bs v\tau)$ does not depend on $\bs x$ and the prior $P_{\bs
  X}(\bs x)$ is a multivariate Gaussian distribution. The integral
with respect to $\bs x$ then becomes, using Eqs.~(A2) and (A10) of
Ref.~\cite{bell1997},
\begin{align} 
\int d\bs x \min\Bk{P_{\bs X}(\bs x),P_{\bs X}(\bs x+\bs v\tau)}
= 
\erfc\bk{\frac{\tau}{\tau_0}},
\label{erfc}
\end{align}
where
\begin{align}
\erfc z &:= \frac{2}{\sqrt{\pi}}\int_z^\infty d\zeta \exp(-\zeta^2),
&
\tau_0 &:= \sqrt{\frac{8}{\bs v^\top\bs \Sigma_0^{-1}\bs v}},
\label{tau0}
\end{align}
and $\bs\Sigma_0 = \expect (\bs X \bs X^\top)-\expect (\bs X) \expect( \bs X^\top)$ is the prior covariance matrix.  A convenient lower
bound on the erfc function is
\begin{align}
\erfc\bk{\frac{\tau}{\tau_0}} &\ge 
\Lambda\bk{\frac{2}{\sqrt{\pi}}\frac{\tau}{\tau_0}},
\label{erfc_bound}
\end{align}
where $\Lambda$ is the triangle function $\Lambda(z) := \max(1-|z|,0)$,
as shown in Fig.~\ref{fig:erfc}.

\section{Multimode quantum optical phase estimation}
\begin{figure}[htbp]
\centerline{\includegraphics[width=0.48\textwidth]{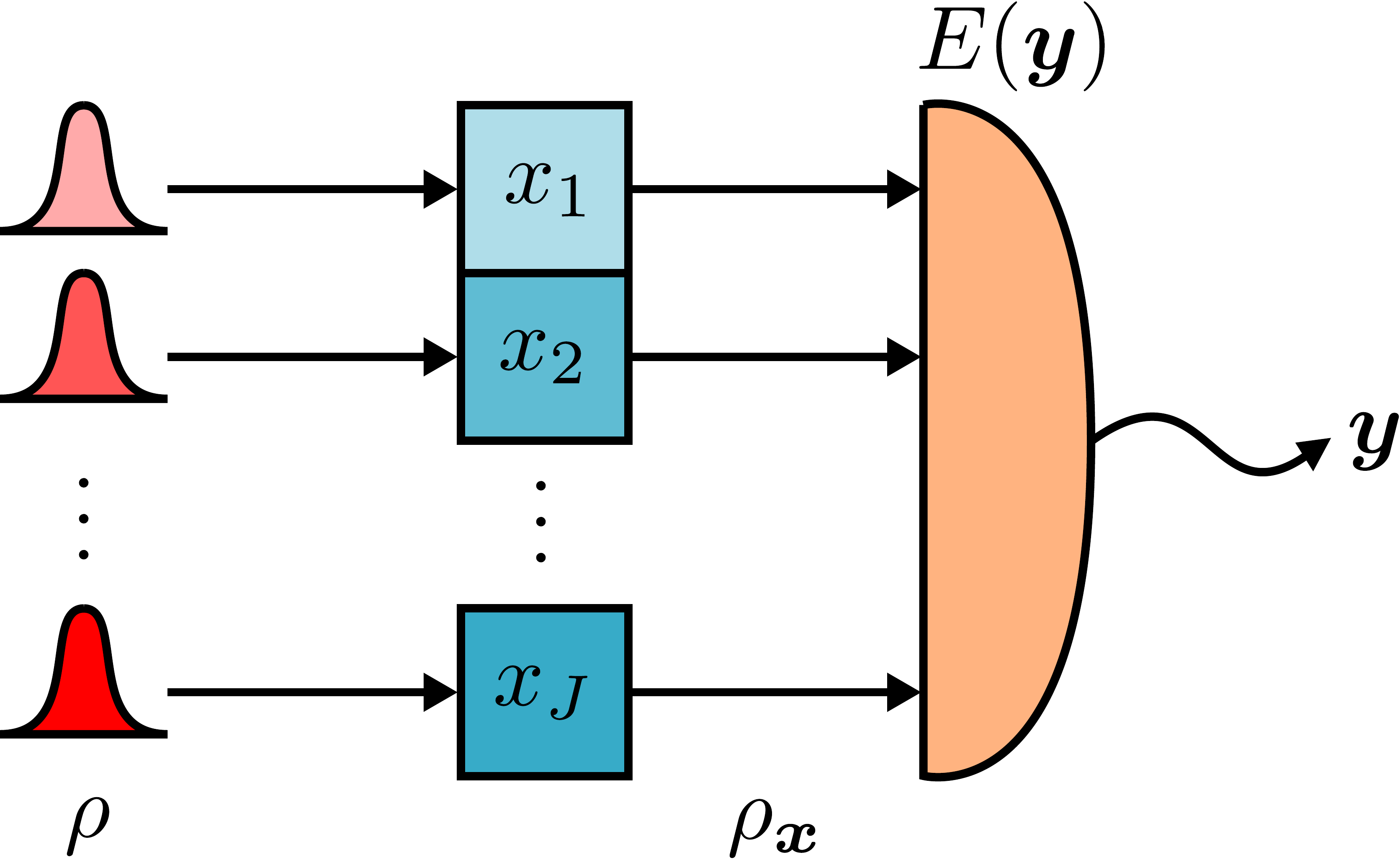}}
\caption{The quantum optical multiparameter phase
estimation problem. The initial state $\rho$ may be entangled between the modes, and each mode passes through a phase shift $x_j$.
The output may be measured via some general joint measurement $E(\bs y)$.}
\label{phasemod}
\end{figure}
We now consider the problem of phase estimation from the measurement
of quantum optical modes, as illustrated in Fig.~\ref{phasemod}.  The
output quantum state is
\begin{align}
\rho_{\bs x} &= \exp\bk{i \bs x^\top\hat{\bs n}}\rho \exp\bk{-i \bs x^\top\hat{\bs n}},
\end{align}
where $\rho$ is the initial quantum state and $\hat{\bs n}$ is a column
vector of photon number operators for the optical modes. 
We use a hat to distinguish the number operator
from other uses of $n$ as an integer.  We will not otherwise use a
hat to indicate operators.
Purifying $\rho_{\bs x}$ to $\ket{\psi(\bs x)}$, and taking the purification of $\rho_{\bs x + \bs v\tau}$ to be
$\exp\bk{i \tau \bs v^\top\hat{\bs n}}\ket{\psi(\bs x)}$,
Uhlmann's theorem \cite{uhlmann} yields a lower bound on $F$
given by \cite{glm_speed}
\begin{align}
&F(\bs x,\bs x+\bs v\tau)
\ge \abs{\Avg{\exp\bk{i\tau\bs v^\top\hat{\bs n}}}}^2
\nn
&= \sum_{\bs n,\bs m}C(\bs n)C(\bs m)
\cos\Bk{\tau \bs v^\top(\bs n-\bs m)},
\end{align}
where we have defined
$\Avg{O}:= \trace\bk{O\rho}$,
and
\begin{align}
C(\bs n) := \avg{\bs n|\rho|\bs n}
\end{align}
is the photon-number distribution of the initial
quantum state, with $\ket{\bs n}$ an eigenstate
of $\hat{\bs n}$.

\begin{figure}[htbp]
\centerline{\includegraphics[width=0.45\textwidth]{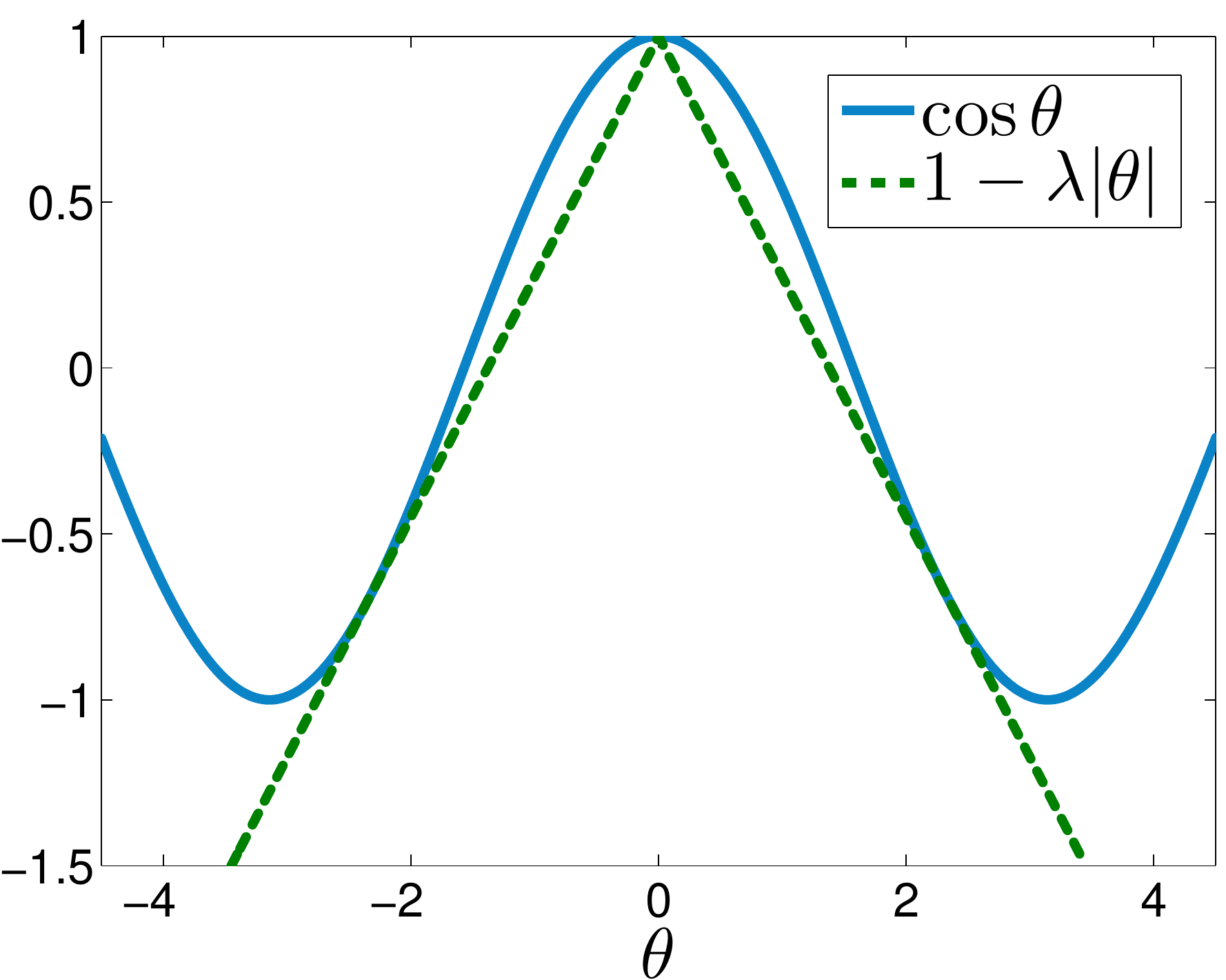}}
\caption{A lower bound on cosine.}
\label{cosine_bound}
\end{figure}

To derive a bound on $F(\bs x,\bs x+\bs v\tau)$
in terms of the average photon numbers, the following
bound on cosine is useful:
\begin{align}
\cos \theta
&\ge 1-\lambda|\theta|,
\end{align}
where $\lambda\approx 0.7246$ is a solution of $\lambda(\pi-\arcsin\lambda)=1+\sqrt{1-\lambda^2}$,
as shown in Fig.~\ref{cosine_bound}. This leads to
\begin{align}
F &\ge \sum_{\bs n,\bs m}C(\bs n)C(\bs m)
\Bk{1-\lambda\tau |\bs v^\top\bk{\bs n-\bs m}|}
\nn
&\ge 
\sum_{\bs n,\bs m}C(\bs n)C(\bs m)
\Bk{1-\lambda\tau\bk{|\bs v^\top\bs n|+|\bs v^\top\bs m|}}
\nn
&\geq 
\sum_{\bs n,\bs m}C(\bs n)C(\bs m)
\Bk{1-\lambda\tau\bk{|\bs v|^\top\bs n+|\bs v|^\top\bs m}}
\nn
&  =   1-2\lambda\tau |\bs v|^\top\Avg{\hat{\bs n}},
\end{align}
where $|\bs v|$ means taking the absolute value of each element of
$\bs v$. Since $0\le F \le 1$, a tighter bound is
\begin{align}
F &\ge \Lambda\bk{\frac{\tau}{\tau_F}},
&
\tau_F &:= \frac{1}{2\lambda |\bs v|^\top\avg{\hat{\bs n}}}.
\label{speed_limit}
\end{align}
A slightly tighter bound may be obtained using the method in
Refs.~\cite{glm_speed,gm_useless}, but the scaling would remain the
same.

Focusing on the mean-square error, putting Eqs.~(\ref{quad_D}),
(\ref{qbzzb4}), (\ref{erfc}), (\ref{erfc_bound}), and
(\ref{speed_limit}) together,
and using $\mathcal V\{ f(\tau) \} \ge f(\tau)$,
\begin{align}
\bs u^\top\bs\Sigma\bs u
&\ge \frac{1}{2}\max_{\bs v: \bs u^\top\bs v = 1}
\int_0^\infty d\tau \, \tau
\Lambda\bk{\frac{2}{\sqrt{\pi}}\frac{\tau}{\tau_0}}
\Lambda\bk{\sqrt{\frac{\tau}{\tau_F}}}
\nn
&= \max_{\bs v: \bs u^\top\bs v=1} Z(\bs v),
\label{qbzzb}
\\
Z(\bs v) &:=
\left\{\begin{array}{ll}
\tau_F^2\bk{\frac{1}{20}-\frac{\tau_F}{21\sqrt{\pi}\tau_0}},
&
\tau_F \le \frac{\sqrt{\pi}\tau_0}{2},
\\
\frac{\pi\tau_0^2}{4}\bk{\frac{1}{12}-\frac{2}{35}
\sqrt{\frac{\sqrt{\pi}\tau_0}{2\tau_F}}},
&
\tau_F > \frac{\sqrt{\pi}\tau_0}{2}.
\end{array}
\right.
\label{Z}
\end{align}
The maximization of $Z(\bs v)$, subject to the constraint $\bs
u^\top\bs v = 1$, gives the tightest bound, but it is difficult to
perform analytically.  In the next section, we shall focus on waveform
estimation and discover that an appropriate choice of $\bs v$, though
suboptimal, can still lead to a reasonably tight bound.

\section{Waveform phase estimation}
\label{sec:est}
\begin{figure}[htbp]
\centerline{\includegraphics[width=0.48\textwidth]{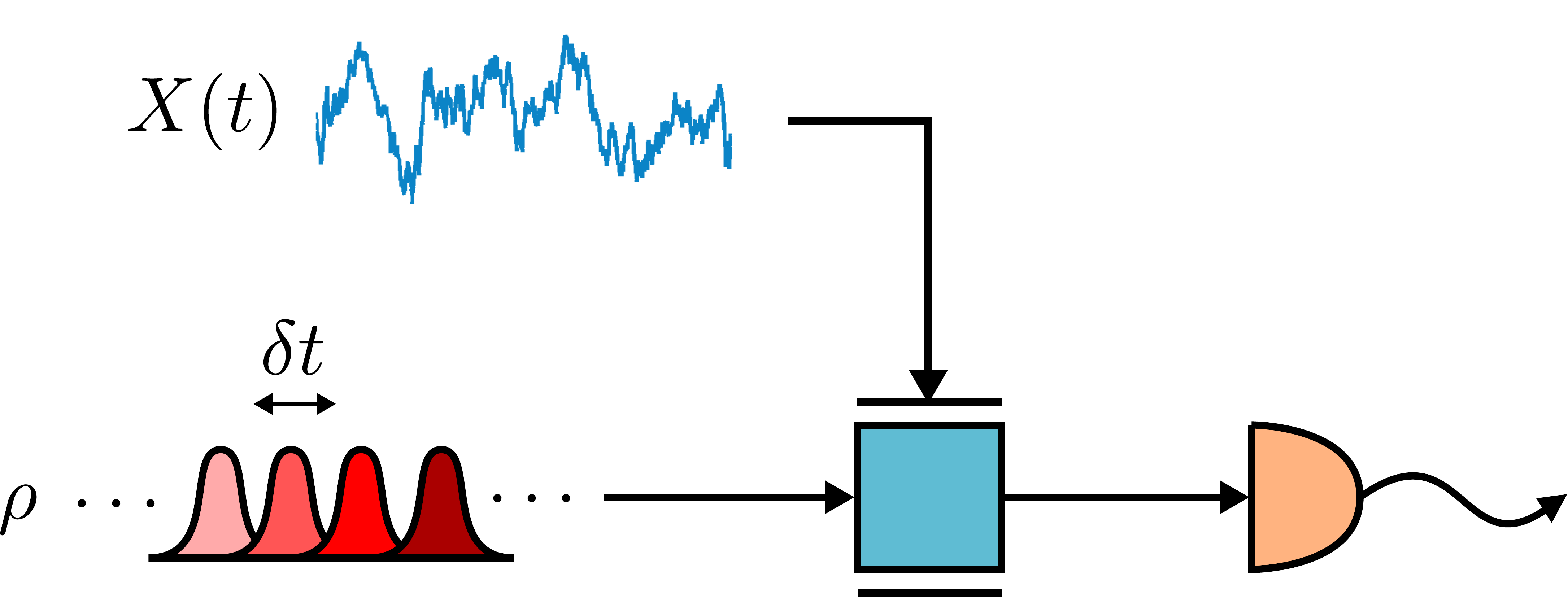}}
\caption{The waveform phase estimation problem.}
\label{waveform}
\end{figure}

We now consider phase modulation that varies in time, as illustrated
in Fig.~\ref{waveform}. Define discrete time as
\begin{align}
t_j &= t_0 + j\delta t,
\end{align}
$j$ being an integer. Each parameter $X_j$ corresponds to a phase at
time $t_j$:
\begin{align}
X_j &= X(t_j),
\end{align}
and each photon-number operator $\hat n_j$ is related to the
photon-flux operator $I(t_j)$ by
\begin{align}
\hat n_j &= I(t_j)\delta t.
\end{align}
Other quantities are redefined as follows:
\begin{align}
u_j &= {u(t_j)/\delta t},
&
v_j &= v(t_j),
\\
\Sigma_{0jk} &= \Sigma_0(t_j,t_k),
&
\Sigma_{jk} &= \Sigma(t_j,t_k).
\end{align}
In the continuous-time limit $\delta t\to 0$,
the mean-square error becomes
\begin{align}
\bs u^\top\bs\Sigma\bs u &\to 
\intall dt \intall dt' \, u(t)\Sigma(t,t')u(t'), 
\end{align}
and the constraint in Eq.~(\ref{qbzzb}) becomes 
\begin{align} \label{conuv}
\bs u^\top \bs v = 1 &\to \intall dt \, u(t)v(t)  = 1.
\end{align}
To evaluate the bound $Z(\bs v)$ given by Eq.~(\ref{Z}) in this limit,
we need to compute $\tau_0$ given by Eq.~(\ref{tau0})
and $\tau_F$ given by Eq.~(\ref{speed_limit}). They depend on
the following:
\begin{align}
\bs v^\top\bs\Sigma_0^{-1}\bs v
&\to \intall dt \intall dt' \, v(t)\Sigma_0^{-1}(t,t')v(t'),
\\
|\bs v|^\top\avg{\hat{\bs n}} &\to \intall dt \, |v(t)|\avg{I(t)},
\end{align}
where the continuous-time inverse $\Sigma_0^{-1}(t,t')$ is defined by
\begin{align}
\intall dt' \, \Sigma_0(t,t')\Sigma_0^{-1}(t',t'') &= \delta(t-t'').
\end{align}
Assume now that the prior statistics of $X(t)$ are stationary. This
means that we can define a prior power spectral density
$\tilde\Sigma_0(\omega)$ such that
\begin{align}
\Sigma_0(t,t') &= \intall \frac{d\omega}{2\pi}\tilde\Sigma_0(\omega)
\exp[i\omega(t-t')],
\end{align}
and the inverse of $\Sigma_0$ is given by
\begin{align}
\Sigma_0^{-1}(t,t') &= \intall \frac{d\omega}{2\pi}\frac{1}{\tilde\Sigma_0(\omega)}
\exp[i\omega(t-t')].
\end{align}
We then obtain
\begin{align}
\bs v^\top\bs\Sigma_0^{-1}\bs v &\to \intall \frac{d\omega}{2\pi} 
\frac{|\tilde v(\omega)|^2}{\tilde\Sigma_0(\omega)},
\label{integral}
\\
\tilde v(\omega) &:=\intall dt \, v(t)\exp(-i\omega t).
\end{align}

We are particularly interested in the estimation error at a particular
time $t_0$, in which case $u(t) = \delta(t-t_0)$ and 
from Eq.~(\ref{conuv}), $v(t_0) = 1$.
We will see below that the choice $\tilde v(\omega) = e^{i\omega t_0}2\pi T\Lambda(T\omega)$, so
\begin{align}
v(t) &= \sinc^2\bk{\frac{t-t_0}{2T}},
\end{align}
where 
\begin{align}
\sinc x &:= \left\{\begin{array}{ll}
(\sin x)/x, & x \neq 0,\\
1, & x = 0, 
\end{array}
\right.
\end{align}
is a convenient one for deriving a lower bound, 
for a suitable choice of characteristic time $T$. It gives 
\begin{align}
\label{fortauF}
|\bs v|^\top\Avg{\hat{\bs n}} &\to \intall dt \, |v(t)| \Avg{I(t)} = 2\pi T \mathcal N(t_0),
\end{align}
where we have defined a weighted average of the flux around $t_0$ by
\begin{align} 
\mathcal N(t_0) := \frac{\intall dt \, |v(t)|\Avg{I(t)}}{\intall dt \, |v(t)|}.
\end{align}

We wish to consider $\tilde\Sigma_0(\omega)$ to be a spectrum with 
power-law scaling as $\kappa^{p-1}/|\omega|^p$ for $\omega$ large.
This scaling is problematic for small $\omega$, because it diverges at $\omega=0$.
To avoid this divergence, we assume \cite{vantrees}
\begin{align}
\label{spect}
\tilde\Sigma_0(\omega) &= \frac{\kappa^{p-1}}{|\omega|^p + \gamma^p},
\end{align}
for some constant $\gamma$. 
For example, $p = 2$ gives the Ornstein-Uhlenbeck process used in
Refs.~\cite{wheatley,yonezawa}. The integral in Eq.~(\ref{integral})
can then be computed analytically, resulting in
\begin{align}
\bs v^\top\bs\Sigma_0^{-1} \bs v &\to 
\frac{8\pi}{p_3 \kappa^{p-1}T^{p-1}} + \frac{4\pi \gamma^p T}{3\kappa^{p-1}}
\approx \frac{8\pi}{p_3 \kappa^{p-1}T^{p-1}},
\nn
p_3 &:= (p+1)(p+2)(p+3),
\label{slope}
\end{align}
where the approximation assumes
\begin{align}
\gamma T \ll \bk{\frac{6}{p_3}}^{1/p},
\label{gammaT}
\end{align}
which will be justified later.
Under this approximation, we will find a bound on the mean-square error that
is independent of $\gamma$.
Alternative choices for removing the singularity at $\omega=0$ yield similar results, 
(see Appendix \ref{appa}).
That is, Eq.~\eqref{slope} depends on the scaling of the spectrum for large $\omega$, 
not on the behavior for small $\omega$. 

The largest $Z(\bs v)$ in Eq.~(\ref{Z}) is obtained by setting
\begin{align}
\tau_F &= \frac{\sqrt{\pi}}{2}\tau_0.
\end{align}
Using Eqs.~\eqref{slope} and \eqref{fortauF}, and recalling the definitions of $\tau_0$ and $\tau_F$ from \eqref{tau0} and \eqref{speed_limit}, respectively, we get
\begin{align}
T &= \Bk{\frac{1}{4\pi^2\lambda^2p_3\kappa^{p-1}\mathcal N^2(t_0)}}^{1/(p+1)}.
\end{align}
Equation \eqref{gammaT} can then be justified in the asymptotic high
$\mathcal N$ limit, because $T$ becomes arbitrarily small. The quantum bound in Eq.~(\ref{qbzzb}) becomes
\begin{align}
\Sigma(t_0,t_0) &\ge \frac{11}{420}\bk{\frac{p_3}{4}}^{2/(p+1)}
\Bk{\frac{\kappa}{4\pi\lambda\mathcal N(t_0)}}^{2(p-1)/(p+1)}.
\end{align}
Rather than considering the error at a single time, we wish to bound
the error averaged over time.  This means bounding
\begin{align}
\bar \Sigma := \lim_{T_{\max}\to\infty} 
\frac 1{2T_{\max}}\int_{-T_{\max}}^{T_{\max}}dt_0 \, \Sigma(t_0,t_0),
\end{align}
in terms of the time-averaged flux,
\begin{align}
\mathcal N := \lim_{T_{\max}\to\infty} \frac 1{2T_{\max}}
\int_{-T_{\max}}^{T_{\max}}dt_0\Avg{I(t_0)}.
\end{align}
It is easy to see that the average of $\mathcal N(t_0)$ is equal to
$\mathcal N$.  Next, because $1/x^{2(p-1)/(p+1)}$ is a convex function
(for $p>1$), the time average of $1/[\mathcal N(t_0)]^{2(p-1)/(p+1)}$
is lower-bounded by $1/\mathcal N^{2(p-1)/(p+1)}$ using Jensen's
inequality.  As a result, we obtain the final result
\begin{align}
\label{scaling}
\bar \Sigma &\ge c_Z \bk{\frac{\kappa}{\mathcal N}}^{2(p-1)/(p+1)}, 
\end{align}
where $c_Z$ is the dimensionless constant 
\begin{align}
c_Z &= \frac{11}{420}\bk{\frac{p_3}{4}}^{2/(p+1)}
\bk{\frac{1}{4\pi\lambda}}^{2(p-1)/(p+1)}. 
\end{align}
That is, we have a lower bound on the time-averaged mean-square error
in terms of the   time-averaged   flux. The 
$(\kappa/\mathcal N)^{2(p-1)/(p+1)}$
scaling was previously proposed in Ref.~\cite{bhw2013} as the 
Heisenberg limit  for a stochastically varying phase with a 
power-law spectrum. However, the proof in that work applies only to a
specific class of Gaussian quantum states. Here, we have
proved the scaling for arbitrary quantum states by introducing  the powerful 
new technique of the quantum 
Bell-Ziv-Zakai bound.

\section{Achieving the optimal scaling}
\label{sec:ach}

\begin{figure}[htbp]
\centerline{\includegraphics[width=0.48\textwidth]{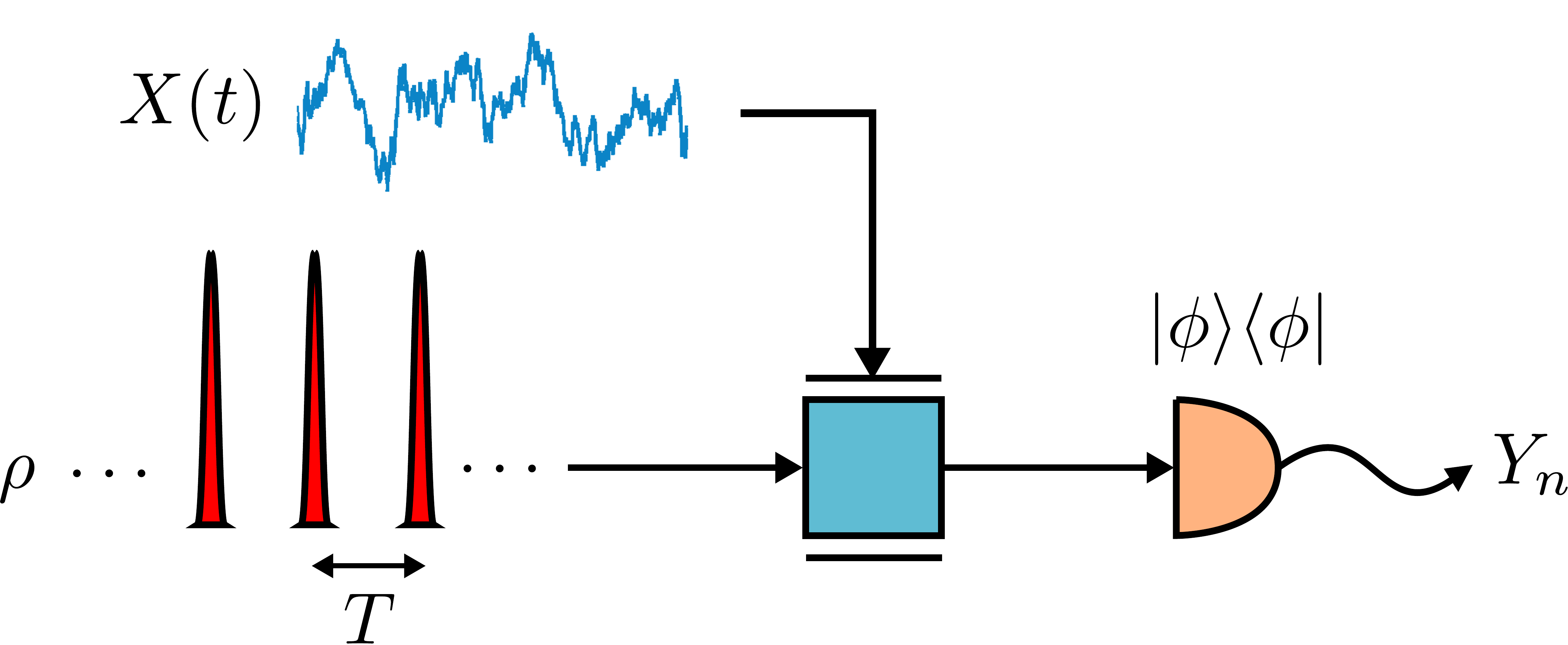}}
\caption{A pulsed phase measurement scheme to achieve the optimal scaling.}
\label{achieve}
\end{figure}

A lower bound is not a Heisenberg limit, and is indeed 
of limited value at all, if it is not close to a realizable
error. Here we demonstrate that the scaling in Eq.~(\ref{scaling}) is
indeed achievable in principle.  Consider an estimation strategy where
the probe field is concentrated into pulses separated by time $T$, as
shown in Fig.~\ref{achieve}. Each pulse is assumed to be so short that
the phase $X(t)$ does not vary during the pulse duration.  The value
that we select for $T$ here will be slightly different than in the
previous section, but the scaling is the same.  With average flux
$\mathcal N$, each pulse can have an average photon number of ${\cal
  N}T$.

We first assume that the phase modulation is weak; viz.,
\begin{align}
\expect \Bk{X^2(t)} &\ll 1.
\label{weakphase}
\end{align}
Using canonical phase measurements and minimum-uncertainty states
within each pulse, the observation $Y_n \in
  (-\pi,\pi]$ at each sampling can be linearized as
\begin{align}
Y_n &\approx X(nT) + \xi_n,
\label{add_noise}
\end{align}
where the moments of the noise random variable $\xi_n$ are
\begin{align}
\label{eq:moments}
\expect\bk{\xi_n|X} \approx 0, \quad
\expect\bk{\xi_n\xi_m|X} \approx \delta_{nm}\frac{(4/27)|z_A|^3}{(\mathcal NT)^2},
\end{align}
with $z_A$ being the first negative root of the Airy function
\cite{bandilla}.  The above moments are exact in the asymptotic limit
of large ${\cal N}T$.  

The condition given by Eq.~(\ref{weakphase}) can be relaxed for large
phase fluctuations by making the canonical phase measurements
adaptive~\cite{wiseman1995}, as shown in Appendix \ref{tracking}. A
rigorous accounting of the error due to phase ambiguity will be
presented in Sec.~\ref{sec:ach2} in the case of a periodic distortion
function. For the remainder of this section we will assume
Eqs.~(\ref{weakphase})--(\ref{eq:moments}) for simplicity.

After all measurements are made, the final estimates can be constructed 
via the Whittaker-Shannon interpolation formula:
\begin{align}
\check X(t) &:= \sum_{n=-\infty}^{\infty} Y_n \sinc\bk{\frac{\pi t}{T}-\pi n}
\nn
&= X_T(t) + \xi(t), \\
\label{XTdef}
X_T(t) &:= \sum_{n=-\infty}^{\infty} X(nT) \sinc\bk{\frac{\pi t}{T}-\pi n},
\\ \label{xidef}
\xi(t) &:=\sum_{n=-\infty}^{\infty} \xi_n \sinc\bk{\frac{\pi t}{T}-\pi n}.
\end{align}
We use this suboptimal interpolation formula rather than optimal
estimation because the error is easier to evaluate. The
mean-square error becomes
\begin{align}
\label{eq:mse}
\expect\Bk{\check X(t)-X(t)}^2
&= \expect\Bk{\xi(t)+X_T(t)-X(t)}^2 \nn
&= \expect\Bk{\xi^2(t)}+\expect\Bk{X_T(t)-X(t)}^2,
\end{align}
which consists of an aliasing error and a measurement error. 
Here we have used the relation 
\begin{align}
\expect\BK{\xi(t)\Bk{X_T(t)-X(t)}} 
&= 0.
\end{align}  
Averaging over time (see Appendix~\ref{appc}), the aliasing error is
\begin{align}
\label{av1}
\frac{1}{T}\int_0^T dt\,\expect\Bk{X_T(t)-X(t)}^2 
&= 
\frac 2\pi \int_{\pi/T}^\infty d\omega\, \tilde\Sigma_0(\omega) 
\nn
&\approx \frac{2(\kappa T)^{p-1}}{\pi^p(p-1)},
\end{align}
which assumes
\begin{align}
\label{gammaT2}
\gamma T \ll \pi,
\end{align}
to be justified later, and the measurement error is
\begin{equation}
\label{av2}
\frac{1}{T}\int_0^T dt \, \expect\Bk{\xi^2(t)}
\approx \frac{(4/27)|z_A|^3}{({\cal N}T)^2}
\end{equation}
via Eq.~(\ref{eq:moments}). The overall error is hence
\begin{equation}
\label{overall}
\bar\Sigma \approx
\frac{2\left(\kappa T\right)^{p-1}}{\pi^p(p-1)}+\frac{(4/27)|z_A|^3}{({\cal N}T)^2}.
\end{equation}
Note that the first term increases with $T$, whereas the second term
decreases with $T$.  This is as we expect, because increasing $T$
means that the phase is sampled less frequently and can vary more
in between samples, but also means that more power is
available to estimate each sample, which reduces the error.

The optimal value of $T$ is
\begin{align}
\label{eq:topt}
T = \left(\frac{(4/27)|z_A|^3\pi^p}{{\cal N}^2\kappa^{p-1}}\right)^{1/(p+1)},
\end{align}
which justifies the assumption in Eq.~(\ref{gammaT2})
in the asymptotic high $\mathcal N$ limit and yields an average variance of
\begin{align}
\label{eq:orgsig}
\bar\Sigma &\approx c_A  (\kappa/{\cal N})^{2(p-1)/(p+1)},
\end{align}
where $c_A > c_Z$ is the dimensionless constant 
\begin{align} 
c_A &= \frac{p+1}{p-1}\bk{4|z_A|^3/27}^{(p-1)/(p+1)}\pi^{-2p/(p+1)}.
\end{align}
Thus the achievable variance has the same scaling with respect to
$\kappa/\cal N$ as that in the lower bound in
Eq.~(\ref{scaling}), but with a larger multiplicative coefficient. 
This demonstrates that the scaling of the lower bound is tight, 
and represents a rigorous Heisenberg limit.

\section{Periodic distortion functions}
\label{sec:per}
Above we have considered phase estimation as an example of the
application of the quantum Bell-Ziv-Zakai bounds.
Phase measurements are intrinsically modulo $2\pi$, because they are
unable to distinguish between phases that differ by multiples of
$2\pi$.  For this reason, phase will typically be taken to be in some
standard region, such as $(-\pi,\pi]$.  Then a phase of
$-\pi+\delta_1$, for some small $\delta_1>0$, can easily be estimated
as $\pi-\delta_2$, for some $\delta_2>0$.  It seems unrealistic to
quantify the error as $\approx 2\pi$, because the phase difference is
small modulo $2\pi$.  For this reason it is better to use periodic
distortion functions for measurements of this type.

\begin{figure}[htbp]
\centerline{\includegraphics[width=0.48\textwidth]{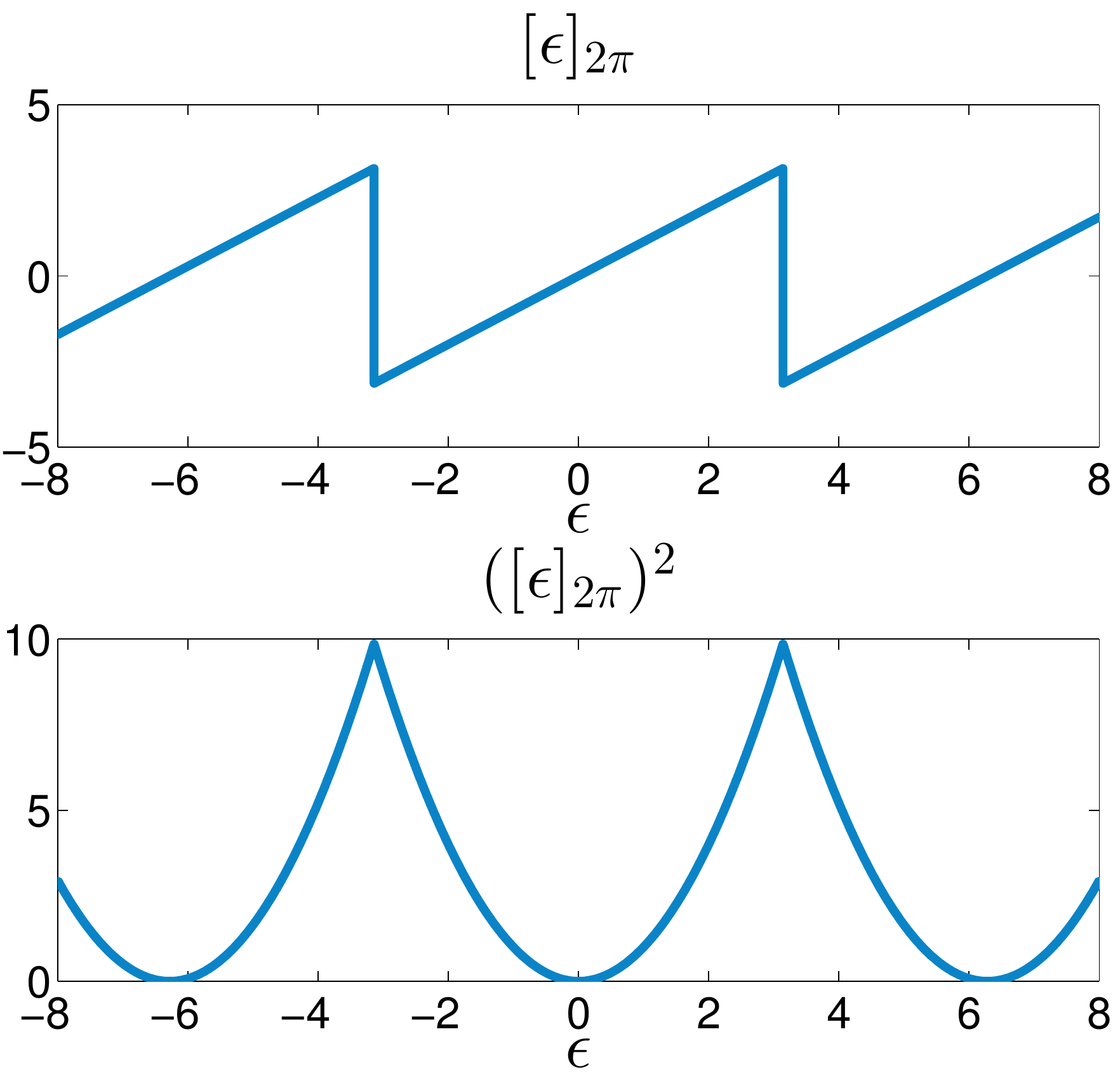}}
\caption{Top: the modulo function $[\epsilon]_{2\pi}$.
Bottom: square of the modulo function.}
\label{modulo}
\end{figure}

In the notation of Ref.~\cite{basu2000}, which we now adopt, 
the distortion function is a vector $\bs D$ with components for each of the parameters $x_j$ to be measured.
For the distortion function to be periodic, it should satisfy
\begin{equation}
D_j(\epsilon_j+2\pi m_j )=D_j(\epsilon_j),
\end{equation}
for any vector of integers $\bs m$.
The distortion function should satisfy most of the conditions used before.
It should be symmetric, have $D_j(0)=0$, and be differentiable and nondecreasing on $[0,\pi)$.
A further condition is that
\begin{equation}
\dot D_j(\epsilon_j) \le \dot D_j(\pi-\epsilon_j),
\end{equation}
for $\pi/2+2\pi m_j \le \epsilon_j \le \pi+2\pi m_j$.
This condition is a technical condition needed for the results of Ref.~\cite{basu2000}.
An example of a periodic distortion function satisfying these conditions is the periodic modification of the mean-square error,
\begin{equation}
D_j(\epsilon_j) = ([\epsilon_j]_{2\pi})^2,
\end{equation}
where the notation 
\begin{align}
[\epsilon]_{2\pi} := \epsilon+2\pi\operatorname{floor}\bk{\frac{1}{2}-\frac{\epsilon}{2\pi}}
\end{align}
denotes the value of $\epsilon$ modulo the interval $(-\pi,\pi]$, as shown in
Fig.~\ref{modulo}.

Using these conditions, it can be shown that (see Eq.~(15) of \cite{basu2000})
\begin{equation}
\expect\Bk{D_j(|\epsilon_j|)} = \frac{1}{2}\int_0^{2\pi} d\tau \, \dot D_j\bk{\frac{\tau}{2}}
\textrm{Pr}\bk{|[\epsilon_j]_{2\pi}| \ge \frac{\tau}{2}}.
\label{D_boundper}
\end{equation}
Note the similarity between this expression for the periodic case and Eq.~\eqref{D_bound} for the non-periodic case.
This expression then gives (see Eq.~(19) of \cite{basu2000})
\begin{align}
&\expect\Bk{D_j(|\epsilon_j|)} \ge \frac{1}{2}\int_0^\pi d\tau \, \dot D_j\bk{\frac{\tau}{2}}
\max_{\bs v:v_j=1}\int_{-\pi}^{\pi} d\bs x \, [P_{\bs X}(\bs{x})
\nonumber\\ &\quad
+ P_{\bs X}([\bs{x}+\bs{v} \tau]_{2\pi})]
P_e(\bs{x},[\bs{x}+\bs{v}\tau]_{2\pi}),
\end{align}
where the bounds on the integral indicate the bounds for each component of $\bs{x}$.

It is easily seen that Property 1 of \cite{bell1997} holds in the periodic case as well, which gives
\begin{align}
&\expect\Bk{D_j(|\epsilon_j|)} \ge \frac{1}{2}\int_0^\pi d\tau \, \dot D_j\bk{\frac{\tau}{2}}
\max_{\bs v:v_j=1}\int_{-\pi}^{\pi} d\bs x \, \min(P_{\bs X}(\bs{x}),
\nonumber\\ &\quad
 P_{\bs X}([\bs{x}+\bs{v} \tau]_{2\pi}))
P_e^{\rm el}(\bs{x},[\bs{x}+\bs{v}\tau]_{2\pi})
\end{align}
Next, using Eq.~\eqref{F_bound},
for the case of quantum measurements we obtain
\begin{align}
\label{qzzb}
\expect\Bk{D_j(|\epsilon_j|)} \ge {}& \frac{1}{2}\int_0^\pi d\tau \, \dot D_j\bk{\frac{\tau}{2}} \max_{\bs v:v_j=1}\int_{-\pi}^{\pi} d\bs x
\nn & \times  \min[P_{\bs X}(\bs{x}),
 P_{\bs X}([\bs{x}+\bs{v} \tau]_{2\pi})]
\nn & \times \Bk{1 - \sqrt{1-F(\bs x,\bs x+\bs v\tau)}}.
\end{align}
Note that $[\bullet]_{2\pi}$ is not needed in the fidelity, because the state is also periodic modulo $2\pi$.

To provide a result for Gaussian variation of $\bs x$ we then encounter a problem.
Gaussian distributions always extend from $-\infty$ to $+\infty$ (even though they exponentially decay),
whereas the variation of $x_j$ is limited to the interval $(-\pi,\pi]$.
Instead the method we use is to take a Gaussian probability distribution $P_G$, and wrap it around $2\pi$.
This would physically correspond to a case where a phase shift is caused by variation in an unbounded quantity (such as the position of a mirror on which the beam is incident \cite{iwasawa}), that has Gaussian statistics.
The probability would then be given by
\begin{equation}
\label{eq:stat}
P_{\bs X}(\bs{x}) = \sum_{\bs{n}} P_G(\bs{x}+2\pi \bs{n}),
\end{equation}
where the sum is over all vectors of integers, $\bs{n}$.
Next, the integral over $\bs{x}$ in Eq.~\eqref{qzzb} can be lower bounded as
\begin{align}
&\int_{-\pi}^{\pi} d\bs{x} \min\Bk{P_{\bs{X}}(\bs{x}), P_{\bs X}([\bs{x}+\bs{v} \tau]_{2\pi})} \nonumber \\
&= \int_{-\pi}^{\pi} d\bs{x} \min\left[\sum_{\bs{n}} P_G(\bs{x}+2\pi \bs{n}), \right. \nonumber \\
& \left. \hspace{10em} \sum_{\bs{m}} P_G(\bs{x}+2\pi \bs{m}  +\bs{v} \tau)\right] \nonumber \\
&\ge \int_{-\pi}^{\pi} d\bs{x} \sum_{\bs{n}} \min\Bk{ P_G(\bs{x}+2\pi \bs{n}), P_G(\bs{x}+2\pi \bs{n}+\bs{v} \tau)} \nonumber \\
&= \int_{-\infty}^{\infty} d\bs{x} \min\Bk{ P_G(\bs{x}), P_G(\bs{x}+\bs{v} \tau)}.
\end{align}

We therefore obtain a result closely analogous to Eq.~\eqref{qbzzb4}: 
\begin{align}
\label{qzzb2}
\expect\Bk{D_j(|\epsilon_j|)} &\ge \frac{1}{2}\int_0^\pi d\tau \, \dot D_j\bk{\frac{\tau}{2}}
\nn &\quad \times \max_{\bs v:v_j=1}\int d\bs x \, \min[P_G(\bs{x}),
 P_G(\bs{x}+\bs{v} \tau)]
\nn &\quad \times \Bk{1 - \sqrt{1-F(\bs x,\bs x+\bs v\tau)}},
\end{align}
where the second integral is over \emph{all} $\bs x$.  On the right-hand
side, the main differences between this expression and
  that in Eq.~\eqref{qbzzb4} are that the integral is up to
  $\tau=\pi$, rather than $\tau\to\infty$, and the valley-filling
  function $\mathcal V$ is not applied here.

Again considering mean-square error for quantum optical phase, Eq.~\eqref{qbzzb} is modified to
\begin{align}
\expect\Bk{D_j(|\epsilon_j|)}
&\ge \frac{1}{2}\max_{\bs v:v_j=1}
\int_0^\pi d\tau \, \tau
\Lambda\bk{\frac{2}{\sqrt{\pi}}\frac{\tau}{\tau_0}}
\Lambda\bk{\sqrt{\frac{\tau}{\tau_F}}}
\nn
&= \max_{\bs v:v_j=1} Z_{\pi}(\bs v),
\label{qbzzb2}
\end{align}
where
\begin{align}
&Z_{\pi}(\bs v) := \nn &
\left\{\begin{array}{ll}
\tau_F^2\bk{\frac{1}{20}-\frac{\tau_F}{21\sqrt{\pi}\tau_0}},
&
\pi \ge \tau_F \le \frac{\sqrt{\pi}\tau_0}{2},
\\
\frac{\pi\tau_0^2}{4}\bk{\frac{1}{12}-\frac{2}{35}
\sqrt{\frac{\sqrt{\pi}\tau_0}{2\tau_F}}},
&
\tau_F > \frac{\sqrt{\pi}\tau_0}{2} \le \pi,
\\
\pi^2 \left(\frac 14 -\frac{\sqrt{\pi}}{3\tau_0}-\frac{\sqrt{\pi}}{5\sqrt{\tau_F}}+\frac{2\pi}{7\tau_0\sqrt{\tau_F}}\right),
& 
\tau_F>\pi<\frac{\sqrt{\pi}\tau_0}{2}.
\end{array}
\right.
\end{align}
The key fact to notice about $Z_\pi$ is that it corresponds to $Z$ when either of $\tau_0$ or $\tau_F$ is small.
In the analysis in Sec.~\ref{sec:est} we took parameters such that both $\tau_0$ or $\tau_F$ are small for large ${\cal N}/\kappa$ (which is the limit we are considering).
Therefore the difference between $F_\pi$ and $F$ has no effect on the bound for the measurements.

The only other difference is that the left-hand side in
Eq.~\eqref{qbzzb2} is the mean-square error, whereas the left-hand
side in Eq.~\eqref{qbzzb} contains the full covariance matrix.  That
is, Eq.~\eqref{qbzzb2} corresponds to taking $u_\ell=\delta_{\ell,j}$,
to give the mean-square error for $x_j$.  However, this is exactly
what is used in Sec.~\ref{sec:est}.  Hence the analysis in
Sec.~\ref{sec:est} continues to hold, and Eq.~\eqref{scaling} is also
a lower bound when the mean-square error modulo $2\pi$ is used.

\section{Achieving the optimal scaling: effect of phase ambiguity}
\label{sec:ach2}
The analysis of the technique for achieving the optimal scaling given in Sec.~\ref{sec:ach} does not fully address the fact that the phase can only be measured modulo $2\pi$.
When tracking a phase, it is possible to resolve this ambiguity from the fact that the variation of the phase is continuous.
Provided the phase does not change too much between successive estimates, and each estimate is reasonably accurate, changes by $2\pi$ can be kept track of.
That is, one can add suitable multiples of $2\pi$ to $Y_n$ to give $\check X_n$, such that $|\check X_{n}-\check X_{n-1}|\le\pi$.
If the initial range of the phase is known, then the error should not exceed $\pi$.

This approach is problematic when the phase can vary arbitrarily far from zero, such as for a Wiener process.
There is a non-zero possibility, however small,
of choosing the wrong interval at any step, and from that point on there will continue to be an error of size $2\pi$ due to this initial error.
This is called a phase-wrap error.
When averaging measurements over an arbitrarily long period of time, the phase error can grow to be arbitrarily large.
For the phase variation we consider, the Fourier spectrum $\tilde\Sigma_0(\omega)$ is bounded for $\omega= 0$, so the prior distribution has a bounded variance for any given time.
This means that the error due to phase-wraps is not unbounded, but it is still problematic.

Here we consider the periodic distortion function, given by the mean-square error modulo $2\pi$.
In this case phase-wrap errors for individual points on their own do not matter, because they do not increase a periodic distortion function.
The problem appears when we consider estimation of the phase \emph{between} the sample points, where the phase is interpolated.
The estimated interpolation error for the Whittaker-Shannon interpolation formula is only accurate if there are no phase-wrap errors.

To simplify the problem, we take $X(t)$ to vary over the entire real line.
Since the error is quantified modulo $2\pi$, the estimation problem is identical to that where $X(t)$ is limited to the region $(-\pi,\pi]$.
There is the additional advantage that the probability distribution is now exactly Gaussian, rather than given by Eq.~\eqref{eq:stat}.

To address the effect of phase-wrap errors on the interpolation, we specify that we consider the mean-square deviation between the interpolated estimate and the actual phase at a given time $t$.
We regard the phase estimate $\check X_{n_t}$ for the sample time nearest $t$ to be in the interval $(-\pi,\pi]$.
This means that the difference between $\check X_{n_t}$ and $X(n_t T)$ will (approximately) be $2\pi K$ for some integer $K$, the number of 
phase wrappings there are between $\check X_n$ and $X(nT)$.
In itself, this difference is unimportant if deviations are only measured modulo $2\pi$.
What is important is that this difference is maintained for the other estimates.
To achieve this, for all other phase estimates $\check X_n$, we add or subtract multiples of $2\pi$ as needed to make the differences between neighboring estimates no more than $\pi$; in particular, $\check X_n-\check X_{n-1}\in (-\pi,\pi]$.
Provided certain conditions are met (discussed below), the difference between $\check X_n$ and $X(nT)$ will be close to $2\pi K$ for all $n$; that is, the \emph{same} multiple of $2\pi$. 
When there are phase-wrap errors, so the difference is close to $2\pi K_n$ where $K_n$ is dependent on $n$, this will introduce error to the interpolation, but this error can be bounded.

Now we make this discussion more rigorous.
First, the noise random variable $\xi_n$ is redefined as
\begin{equation}
\xi_n := [Y_n-X(nT)]_{2\pi}.
\end{equation}
With this definition, the moments given in Eq.~\eqref{eq:moments} are correct.
We can give $\check X_n$ as
\begin{equation}
\label{eq:kn}
\check X_n = X(nT) + \xi_n + 2\pi K_n,
\end{equation}
Recall that $Y_n$ is the measurement result in the interval $(-\pi,\pi]$, which is then adjusted to $\check X_n$ by adding or subtracting multiples of $2\pi$.

The interpolated values $\check X(t)$ can be expressed as
\begin{equation}
\label{ester}
\check X(t) = X_T(t) + \xi(t) + 2\pi K(t),
\end{equation}
where $X_T(t)$ and $\xi(t)$ are defined as in Eqs.~\eqref{XTdef} and \eqref{xidef}, and
\begin{equation}
 K(t) := \sum_{n=-\infty}^{\infty} K_n \sinc\bk{\frac{\pi t}{T}-\pi n}.
\end{equation}
This shows why we want the difference between the estimates and values to remain the same multiple of $2\pi$.
If they do, then $K(t)$ simply becomes the constant $K$.
On the other hand, if $K_n$ varies with $n$, then $K(t)$ will not be an integer, and the extra term $2\pi K(t)$ in Eq.~\eqref{ester} will give an increased error modulo $2\pi$.

The analysis of the error in Eq.~\eqref{eq:mse} needs to be performed modulo $2\pi$.
First, define
\begin{equation}
\Delta(t) := [\xi(t)+X_T(t)-X(t)]_{2\pi}.
\end{equation}
In terms of this quantity we obtain
\begin{align}
\label{eq:nonin}
&\expect\Bk{[\check X(t)-X(t)]_{2\pi}^2} = \expect\Bk{\left[ 2\pi K(t) + \Delta(t)\right]_{2\pi}^2} \nn
&\le \expect\Bk{ \left\{[2\pi K(t)]_{2\pi} + \Delta(t)\right\}^2} \nn
&=\expect\Bk{ [2\pi K(t)]_{2\pi}^2)}+\expect\Bk{\Delta^2(t)} + 2\expect\Bk{ \Delta(t)[2\pi K(t)]_{2\pi} } \nn
&\le \left\{ \sqrt{\expect\Bk{ [2\pi K(t)]_{2\pi}^2}} + \sqrt{\expect\Bk{\Delta^2(t)}} \right\}^2.
\end{align}

The time-averaged value of $\expect\Bk{\Delta^2(t)}$ is exactly what was obtained as in Sec.~\ref{sec:ach}, with the result given in Eq.~\eqref{overall}. Note also that we can 
upper bound the time-averaged value of $\expect\Bk{[\check X(t)-X(t)]_{2\pi}^2}$ using the time averaged values of
$\expect\Bk{\Delta^2(t)}$ and $\expect\Bk{[2\pi K(t)]_{2\pi}^2}$.
That is because $\expect\Bk{[\check X(t)-x(t)]_{2\pi}^2}$ is a concave function of these quantities.
The remaining task is therefore to find the time-averaged value of $\expect\Bk{[2\pi K(t)]_{2\pi}^2}$.

To achieve this, we need to bound the probabilities of phase-wrap errors.
If $t$ is a sample time so $t=n_t T$, then $K(t)=K_{n}$, and $[2\pi K(t)]_{2\pi}$ is equal to zero.
Therefore, in the remainder of this discussion, we will assume that $t$ is not a sample time.
Let $L_n$ be the multiple of $2\pi$ that we have added to the measurement result $Y_n$ to give $\check X_n$; that is
\begin{equation}
\label{eq:ln}
\check X_n = Y_n + 2\pi L_n.
\end{equation}
We wish to consider the error in interpolating at the given time $t$.
Without loss of generality, this time can be taken to be in the interval $(0,T/2]$.
This is because there is translation symmetry and time-reversal symmetry.
One can simply translate the time by a multiple of $T$, and change the sample numbering such that $n=0$ or $n=1$ corresponds to the closest sample time.
That would yield $t\in(0,T)$.
Then, if $t\in (T/2,T)$, one can simply reverse all times about $T/2$, so $t\in(0,T/2]$.

We can select $n_t={\rm round}(t/T)$ with the ``round half down'' convention, so if $t$ is equidistant between two sample times, we take the smaller sample time.
In that case, for $t\in(0,T/2]$, $n_t=0$.
We are then starting with $\check X_0$ taken to be in the interval $(-\pi,\pi]$, so $\check X_0=Y_0$ and $L_0=0$. 
Then all other values of $L_n$ are selected such that $|\check X_{n+1}-\check X_n|\le \pi$.
The goal of this is to ensure that the values of $K_n$ are equal (or at least close) to $K_0$.
Using Eqs.~\eqref{eq:kn} and \eqref{eq:ln}, we obtain
\begin{align}
&2\pi |K_{n+1}-K_{n}| \nn
&= |\check X_{n+1}-\check X_n+\xi_n-\xi_{n+1} + X(nT)- X((n+1)T)| \nn
& \le |\check X_{n+1}-\check X_n|+|\xi_n-\xi_{n+1}| + |X(nT)- X((n+1)T)| \nn
& \le \pi + |\xi_n-\xi_{n+1}| + |X(nT)- X((n+1)T)|.
\end{align}
In the last line we have used the fact that the values of $L_n$ have been chosen such that $|\check X_{n+1}-\check X_n|\le \pi$.
Now, if it is the case that $|\xi_n-\xi_{n+1}|  + |X(nT)- X((n+1)T)|$ is \emph{less} than $\pi$, then $2\pi |K_{n+1}-K_{n}|<2\pi$.
Because $K_n$ takes integer values, this inequality implies that $K_{n+1}=K_{n}$.

In the following, we will wish to ensure that $|\xi_n-\xi_{n+1}|<\pi/2$, and $|X(nT)- X((n+1)T)|< \pi/2$.
Note that $|\xi_n|\le \pi$, so we must always obtain $|\xi_n-\xi_{n+1}|\le 2\pi$.
Below we will show that the probability of $|X(nT)- X((n+1)T)|\ge \pi/2$ is negligible.
Given that $|X(nT)- X((n+1)T)|< \pi/2$ and $|\xi_n-\xi_{n+1}|\le 2\pi$, we must have
$2\pi |K_{n+1}-K_{n}| < 4\pi$.
This ensures that $K_n$ cannot change by more than $1$; that is, we do not have more than $1$ phase-wrap error at a time.

To consider the effect of a phase-wrap error, let $Z_{n}:=K_{n}-K_{n-1}$, which can take values $Z_n\in\{0,-1,+1\}$ with non-negligible probability (since the probability of multiple phase-wrap errors is insignificant).
Then we obtain, for positive $n$,
\begin{equation}
K_n = K_0 + \sum_{m=1}^n Z_m.
\end{equation}
Similarly, for negative $n$,
\begin{equation}
K_n = K_0 - \sum_{m=n+1}^{0} Z_m.
\end{equation}
Therefore we can write $K(t)$ as
\begin{align}
\label{eq:digam}
K(t) &= K_0 + \sum_{n=1}^{\infty} \sum_{m=1}^{n} Z_m \sinc\left(\frac{\pi t}{T}-\pi n\right) \nn
&\quad+\sum_{n=-\infty}^{-1} \sum_{m=n+1}^{0} Z_m \sinc\left(\frac{\pi t}{T}-\pi n\right) \nn
&= K_0 +\sum_{m=1}^{\infty} Z_m \sum_{n=m}^{\infty}  \sinc\left(\frac{\pi t}{T}-\pi n\right) \nn
&\quad+\sum_{m=-\infty}^{0} Z_m \sum_{n=-\infty}^{m-1} \sinc\left(\frac{\pi t}{T}-\pi n\right) \nn
&= K_0 + \frac 1{2\pi} \sum_{m=-\infty}^{\infty} Z_m a_m,
\end{align}
where (see Appendix \ref{digamma})
\begin{align}
&a_m := (-1)^m \sin\left(\frac{\pi t}{T}\right)  \nn
&\times\begin{cases}
\dig \left(\frac{m}{2}-\frac{t}{2T}\right)- \dig \left(\frac 12+\frac{m}{2}-\frac{t}{2T}\right),    & m>0\\
\dig \left(\frac 12-\frac m2+\frac t{2T}\right)- \dig \left(1-\frac m2+\frac t{2T}\right),    & m\le 0,
\end{cases}
\end{align}
where $\dig$ is the digamma function $ \dig (x)=\Gamma'(x)/\Gamma(x)$.

Due to symmetry, $\expect\Bk{Z_m} =0$.
Also, note that $\expect\Bk{ Z_n Z_m}\approx 0$ for $|n-m|>1$.
This is because the errors in the phase estimates are independent, so the probability of $|\xi_n-\xi_{n+1}|$ exceeding $\pi/2$ is independent of the probability of $|\xi_m-\xi_{m+1}|$ exceeding $\pi/2$.
This means that the only way that $Z_n$ can be correlated with $Z_m$ is through correlations in the variation of $X$.
That is, the probability of $|X(nT)- X((n+1)T)|$ exceeding $\pi/2$ is correlated with that for $|X(mT)- X((m+1)T)|$.
However, because the probability for this is negligible, the overall correlations are negligible.
For a rigorous proof that $\expect\Bk{ Z_n Z_m}$ can be neglected, see Appendix \ref{corrbou}.

As a result, we can bound $\expect\Bk{[2\pi K(t)]_{2\pi}^2}$ as follows:
\begin{align}
\label{longapprox}
&\expect\Bk{ [2\pi K(t)]_{2\pi}^2} \nn &= \expect\Bk{ [2\pi (K(t)-K_0)]_{2\pi}^2} \nn
&\le \expect\Bk{ [2\pi (K(t)-K_0)]^2} \nn
&= \expect\Bk{ \left( \sum_{m=-\infty}^{\infty} Z_m a_m \right)^2 } \nn
&\approx \sum_{m=-\infty}^{\infty} \expect\Bk{ Z_m^2 } a_m^2 + 2\sum_{m=-\infty}^{\infty} \expect\Bk{ Z_m Z_{m+1} } a_m a_{m+1} \nn
&\lapprox p_{\rm err} \left[ \sum_{m=-\infty}^{\infty} a_m^2 + 2\sum_{m=-\infty}^{\infty} |a_m a_{m+1}| \right],
\end{align}
where $\lapprox$ indicates that terms exponentially small in ${\cal N}/\kappa$ have been omitted, and $p_{\rm err}$ is the probability of a phase wrap error at each step.
Because $Z_m$ is limited to $\{0,-1,+1\}$ with high probability, $\expect\Bk{ Z_m^2 } \approx p_{\rm err}$ (see Appendix \ref{corrbou}).
In addition, $|\expect\Bk{ Z_m Z_{m+1} }|$ cannot exceed $\expect\Bk{ Z_m^2 }$, which gives the inequality in the last line.

Numerical calculation of the quantity in square brackets on the last line gives the maximum value for $t/T=1/2$ as $0.68169\approx 1-1/\pi$.
Therefore we have
\begin{equation}
\label{eq:wrapbnd}
\expect\Bk{ [2\pi K(t)]_{2\pi}^2} \lapprox p_{\rm err} (1-1/\pi).
\end{equation}

The next task is to bound $p_{\rm err}$.
We first consider the difference between $X(nT)$ and $X((n-1)T)$.
It turns out that $\expect\Bk{[X(t)-X(t')]^2}$ can be bounded as a polynomial in $\kappa/{\cal N}$ (see Appendix \ref{Xbou}).
As we are considering the scaling with small $\kappa/{\cal N}$, and the statistics of the variation are Gaussian, the probability of the difference between $X(nT)$ and $X((n-1)T)$ being larger than $\pi/2$ is exponentially small.
Because we only consider results polynomial in $\kappa/{\cal N}$, this exponentially small probability can be ignored without altering the asymptotic scaling.

Next we consider the probability of $|\xi_n-\xi_{n-1}|$ exceeding $\pi/2$.
The variance in these estimates scales as $(4/27)|z_A|^3/({\cal N}T)^2$.
Because the error in these estimates is independent, the variance in their difference is $\sim (8/27)|z_A|^3/({\cal N}T)^2$.
Using Markov's inequality, the probability of $|\xi_n-\xi_{n-1}|$ being larger than $\pi/2$ cannot be larger than
\begin{equation}
\frac{4}{\pi^2}\frac{(8/27)|z_A|^3}{({\cal N}T)^2}.
\end{equation}
As this is the dominant term in the probability of a phase error, we have
\begin{equation}
p_{\rm err} \lapprox \frac{8}{\pi^2}\frac{(4/27)|z_A|^3}{({\cal N}T)^2}.
\end{equation}

Using this together with Eq.~\eqref{eq:wrapbnd} gives
\begin{equation}
\expect\Bk{ [2\pi K (t)]_{2\pi}^2} \lapprox \frac{8(1-1/\pi)}{\pi^2}\frac{(4/27)|z_A|^3}{({\cal N}T)^2}.
\end{equation}
This value is for the worst case value of $t$ (i.e., midway between two sample points), so averaging over $t$ can only give smaller values.

Rather than rederiving an optimal value of $T$, we can simply use Eq.~\eqref{eq:topt}.
The scaling for the average variance given in Eq.~\eqref{eq:orgsig} corresponds to the time average of $\expect\Bk{ \Delta^2(t)}$ with $T$ given as in Eq.~\eqref{eq:topt}.
Denoting this quantity by $\bar\Sigma_0$, and the time averaged value of $\expect\Bk{ [2\pi K (t)]_{2\pi}^2}$ by $\bar\Sigma_{\rm wrap}$,
using Eq.~\eqref{eq:nonin} then gives
\begin{align}
&\bar\Sigma \lapprox \left( \sqrt{\bar\Sigma_{\rm wrap}} + \sqrt{\bar\Sigma_0} \right)^2 \nn
&\lapprox \left(\sqrt{\frac{p+1}{p-1}}+\sqrt{\frac{8(1-1/\pi)}{\pi^2}}\right)^2 \frac{[(4/27)|z_A|^3]^{(p-1)/(p+1)}}{\pi^{2p/(p+1)} }\nn
&\quad \times (\kappa/{\cal N})^{2(p-1)/(p+1)}.
\end{align}
Thus we find that, when we fully take account of phase-wrap errors, we still obtain
\begin{align}
\bar\Sigma  = O\left((\kappa/{\cal N})^{2(p-1)/(p+1)}\right).
\end{align}

\section{Conclusions}
While fundamental quantum limits to accuracy of measurement of single
quantities are well-known, deriving fundamental limits becomes very
challenging when there is prior information and correlations between
the quantities to be measured.  A particularly important example of
this is in phase estimation, where the phase at any
time is correlated with the phase at earlier and later times.  This
task is needed, for example, in gravitational wave
astronomy.

Here we have proven quantum forms of the
Bell-Ziv-Zakai bounds for multiparameter estimation.
One of the bounds enables us to bound the accuracy
possible when measuring a phase with stationary Gaussian prior
statistics and a power-law spectrum.  We have thereby been able to
prove that the scaling bound found in Ref.~\cite{bhw2013}, for quantum states
having time-symmetric stationary Gaussian statistics for the field
quadratures, in fact holds for all possible quantum states.

Moreover, we have shown here analytically that the lower bound we have derived is always 
 achievable, up to a constant factor. Specifically, it is possible to achieve it 
 by sampling with regularly timed sequence of pulses,  each of which is measured 
 by a canonical phase measurement, and with interpolation of the phase between those times.
This bound can therefore be regarded as analogous to the Heisenberg limit for measurement of a single constant phase. 
We have also provided bounds for periodic distortion functions.
An example of this is measurement of phase modulo $2\pi$, so the mean-square error is evaluated modulo $2\pi$.
We find that the bounds we derive for the nonperiodic case hold almost unchanged.

For the future, it is still an open question as to whether our phase
estimation bound could be achieved more simply, for example using
continuous (rather than pulsed) Gaussian field states with suitable
correlations, and using homodyne detection (perhaps
adaptive~\cite{wiseman1995}) rather than assuming canonical phase
measurements. We also note that while Gaussian correlations were assumed for the applied phase shift ${\bf X}$ (e.g., in Eq.~(\ref{erfc})), our method readily generalizes to yield estimation bounds for non-Gaussian correlations. More generally, there are many other multiparameter
estimation tasks, in which there are prior constraints on the
correlations, for which our quantum Bell-Ziv-Zakai bounds could reveal
the ultimate achievable limits.

\section*{Acknowledgments}
Discussions with Ranjith Nair are gratefully acknowledged.  DWB is
supported by ARC grant FT100100761.  MT is supported by the Singapore
National Research Foundation under NRF Grant No.~NRF-NRFF2011-07.
MJWH and HMW are supported by the ARC Centre of Excellence CE110001027.

\appendix
\section{Asymptotic scaling for spectra with power-law tail}
\label{appa}
More generally, the prior power spectral density
$\tilde\Sigma_0(\omega)$ may scale as $\kappa^{p-1}/|\omega|^p$ for
$|\omega|\to\infty$.  Making this concept rigorous, we assume that
there exist constants $w_0$ and $G$ such that
\begin{align}
\tilde\Sigma_0(\omega) &\ge \left\{
\begin{array}{ll}
G/\kappa, & |\omega| < \kappa w_0,\\
\kappa^{p-1}/|\omega|^p, & |\omega| \ge \kappa w_0.
\end{array}
\right.
\end{align}

Then we obtain
\begin{align}
\bs v^\top\bs\Sigma_0^{-1} \bs v &\to 2\pi T^2\int_{-1/T}^{1/T}  d\omega \, \frac{(1-|\omega| T)^2}{\tilde\Sigma_0(\omega)} \nn
&\le 4\pi T^2\kappa \int_0^{\kappa w_0} d\omega \frac{(1-\omega T)^2}{G} \nn &\quad+
4\pi T^2 \int_0^{1/T} d\omega \frac{\omega^p(1-\omega T)^2}{\kappa^{p-1}} \nn
&\le \frac {4\pi T^2 \kappa^2 w_0}{G} +\frac{8\pi}{p_3 \kappa^{p-1} T^{p-1}}.
\end{align}
The first term is negligible provided $\kappa T$ is small.
With the expression we take for $T$, $\kappa T\propto
[\kappa/\mathcal N(t_0)]^{2/(p+1)}$.  Because we consider scaling with large
$\mathcal N(t_0)/\kappa$, the first term is negligible and we again obtain the result in the main text.

\section{Canonical phase-locked loop}
\label{tracking}
The accuracy of our linear model in Sec.~\ref{sec:ach} relies on the
assumption of weak phase modulation. For large phase fluctuations, we
can borrow from the phase-locked loop concept
\cite{wheatley,yonezawa,iwasawa,wiseman1995,armen,berry2002,berry2006,tsl2008,tsl2009,vantrees2}
and modulate each pulse by an adaptive phase $-\tilde X(nT)$ before
the canonical phase measurement, where $\tilde X(nT)$ is a causal
estimate of $X(nT)$ extrapolated from previous observations
$\{Y_{n-1},Y_{n-2},\dots\}$.  Provided that $\tilde X(nT)$ tracks
$X(nT)$ closely; viz.,
\begin{align}
\expect\Bk{[X(nT)-\tilde X(nT)]^2} &\ll 1,
\label{small_causal}
\end{align}
the net phase modulation $X(nT) - \tilde X(nT)$ will be small, and
$Y_n \in (-\pi,\pi]$ can be linearized as
\begin{align}
Y_n &\approx X(nT) - \tilde X(nT) + \xi_n.
\end{align}
The requirement of small causal error according to
Eq.~(\ref{small_causal}) is now less stringent than
Eq.~(\ref{weakphase}). To evaluate the causal error analytically, we
approximate the discrete observations $Y_n$ as a continuous-time
signal given by
\begin{align}
Y(t) &\approx X(t) -\tilde X(t) + \xi(t),
\\
\expect\Bk{\xi(t)\xi(t')|X} &= R\delta(t-t'),
\\
R &:= \frac{(4/27)|z_A|^3}{{\cal N}^2T}.
\end{align}
The continuous approximation is accurate in the high $\mathcal N$
limit because the measurement period $T$ in Eq.~(\ref{eq:topt}) can be
made arbitrarily small in the limit.  The minimum causal error at
steady state is then given by the Yovits-Jackson formula
\cite{vantrees}:
\begin{align}
\expect\Bk{[X(t)-\tilde X(t)]^2} &\approx R 
\intall \frac{d\omega}{2\pi} \ln \Bk{1+\frac{\tilde\Sigma_0(\omega)}{R}}.
\end{align}
Since the error decreases with decreasing $R$ and increasing $\cal N$,
the phase tracking can be made arbitrarily accurate in the high $\cal
N$ limit. These considerations are similar to the principles of a
homodyne phase-locked loop
\cite{vantrees2,berry2002,berry2006,tsl2008,tsl2009}, except that here
we assume canonical phase measurements to avoid photon-number
fluctuations. 

In the long-time limit, phase-wrap errors, no matter how rare, can
still occur, making the estimate diverge from the true waveform by
multiples of $2\pi$. Just like the classical phase
  modulation system, it can be expected that this divergence will be
  eliminated by adding a DC notch filter to the output
\cite{vantrees2}.

\section{Time averages}
\label{appc}
Here we show how to take the time averages \eqref{av1} and \eqref{av2} given in the main text.
Each average is taken because the error will depend on how far $t$ is from the nearest sampling point.
Because the distribution is otherwise time invariant, we need only average over the interval $[0,T]$.

Here we take $S(t):=\Sigma_0(t,0)$ and $f:=1/T$.
Note that, due to stationary statistics, $\Sigma_0(t_1,t_2)$ depends only on $t_1-t_2$.
Evaluating Eq.~\eqref{av1} gives
\begin{widetext}
\begin{align}
&\frac 1{T}\int_{0}^{T} dt \, \expect\Bk{X_T(t)-X(t)}^2 
= \frac 1{T}\int_{0}^{T} dt\,  
\left[ \sum_{n=-\infty}^{\infty} \sum_{m=-\infty}^{\infty} \expect\Bk{X(nT)X(mT)} \sinc(\pi(ft-n))\sinc(\pi(ft-m)) \right.
\nn
&  \left. \quad +\expect\Bk{X(t)X(t)} -2\sum_{n=-\infty}^{\infty} \expect\Bk{ X(nT)X(t)}  \sinc(\pi(ft-n)) \right] \nn
&= \frac 1{T}\int_{0}^{T} dt\,  \left[ \sum_{m=-\infty}^{\infty} \sum_{n=-\infty}^{\infty} S(nT) \sinc(\pi(ft-n-m)) \sinc(\pi(ft-m)) + S(0) -2\sum_{n=-\infty}^{\infty} S(t-nT) \, \sinc(\pi(ft-n)) \right] \nn
&= S(0) + \frac 1{T}\int_{-\infty}^{\infty} dt\,  \left[ \sum_{n=-\infty}^{\infty} S(nT) \sinc(\pi(ft-n))\sinc(\pi ft)  -2S(t) \, \sinc(\pi ft) \vphantom{ \sum_{n=-\infty}^{\infty}} \right] \nn
&= S(0) + \sum_{n=-\infty}^{\infty} S(nT) \, \sinc(\pi n) -\frac 2{T}\int_{-\infty}^{\infty} dt\,  S(t) \, \sinc(\pi ft) \nn
&= 2S(0) -\frac 2{T}\int_{-\infty}^{\infty} dt\, S(t) \, \sinc(\pi ft) = 2S(0) -\frac 1{\pi}\int_{-\pi f}^{\pi f} d\omega \, \int_{-\infty}^{\infty} dt\,   S(t) \,  e^{-i\omega t} \nn
&= \frac 1{\pi}\int_{-\infty}^{\infty} d\omega \, \tilde\Sigma_0(\omega)-\frac 1{\pi}\int_{-\pi f}^{\pi f} d\omega \, \tilde\Sigma_0(\omega) = \frac{2}{\pi}\int_{\pi f}^{\infty} d\omega \, \tilde\Sigma_0(\omega).
\end{align}
\end{widetext}
Evaluating the integral for   $\tilde\Sigma_0(\omega)=\kappa^{p-1}/(|\omega|^p+\gamma^p)$   gives
\begin{align}
\frac{2}{\pi}\int_{\pi f}^{\infty} d\omega \, \tilde\Sigma_0(\omega)
&= \frac{2}{\pi}\int_{\pi f}^{\infty} d\omega \, \frac{\kappa^{p-1}}{\omega^p+\gamma^p}\nn
&\approx \frac{2\left(\kappa T\right)^{p-1}}{\pi^p(p-1)},
\end{align}
assuming $\pi f = \pi/T \gg \gamma$.

Next, evaluating the average in Eq.~\eqref{av2} gives
\begin{align}
&\frac{1}{T}\int_0^T dt \, \expect\Bk{ \xi^2(t)} \nn
&= \frac{1}{T}\int_0^T dt \sum_{n=-\infty}^{\infty} \expect\Bk{ \xi_n^2 } \sinc^2(\pi(ft-n)) \nn
&\sim \frac{1}{T}\int_0^T dt \sum_{n=-\infty}^{\infty} \frac{(4/27)|z_A|^2}{({\cal N}T)^2} 
\sinc^2(\pi(ft-n)) \nn
&= \frac{(4/27)|z_A|^2}{({\cal N}T)^2} \frac{1}{T}\int_{-\infty}^{\infty} dt \sinc^2(\pi ft) \nn
&= \frac{(4/27)|z_A|^2}{({\cal N}T)^2}.
\end{align}

\section{Proof of formula for $a_m$}
\label{digamma}
To prove the last line of Eq.~\eqref{eq:digam}
for $m>0$, consider the semi-infinite sum
\begin{equation}
s_m(x):=\sum_{n=m}^{\infty}f(x-n\pi) =s_0(x-m\pi),
\end{equation}
for some function $f(x)$.  Noting that $s_0(x)=f(x)+s_0(x-\pi)$, it follows that $s_0(x)=g(x)+p(x)$ where $g(x)$ is any solution of the recurrence relation $g(x)-g(x-\pi)=f(x)$ and $p(x)$ is some periodic function with period $\pi$.  Now, using the known recurrence relation $\dig(z+1)=\dig(z)+1/z$, it follows that $g(x):=\sin(x) \left[\dig(-x/2\pi)-\dig(1/2-x/2\pi) \right]$ satisfies $g(x)-g(x-\pi) = 2\pi \sinc x$. Hence,
\begin{equation}
 \sum_{n=m}^\infty \sinc (x-n\pi) = (2\pi)^{-1} g(x-m\pi) + p(x)
\end{equation}
for some periodic function $p(x)$ with period $\pi$.
Both the sum and $g(x-m\pi)$ vanish in the limit $m\rightarrow\infty$, so $p(x)=0$.
Taking $x=\pi t/T$ and using $\sin(\pi t/T-m\pi)=(-1)^m \sin(\pi t/T)$ proves the formula for $m>0$.
The proof for $m\leq 0$ is similar.

\section{Bounding change in $X$}
\label{Xbou}
Here we show how to bound $\expect\Bk{[X(t)-X(t')]^2}$.
In general, using only the property that $\tilde\Sigma_0(\omega)$ is an even function,
\begin{align}
& \expect\Bk{[X(t)-X(t')]^2} = 2[\Sigma_0(t,t)-\Sigma_0(t,t')] \nonumber \\
&= \int_{-\infty}^{\infty} \frac{d\omega}{\pi} \tilde\Sigma_0(\omega) \{1-\exp[i\omega(t-t')]\} \nonumber \\
&= \int_{-\infty}^{\infty} \frac{d\omega}{\pi} \tilde\Sigma_0(\omega) \{1-\cos[\omega(t-t')]\}.
\end{align}
To bound the variance, rather than using the spectrum in the form \eqref{spect}, we use upper bounds.
In the case for $1<p<3$, it is convenient to use the upper bound $\tilde\Sigma_0(\omega)<\kappa^{p-1}/|\omega|^p$, which gives for $p\ne 2$
\begin{align}
\label{p1to3}
& \expect\Bk{[X(t)-X(t')]^2} \nn
&< (\kappa|t-t'|)^{p-1}\frac{-2\Gamma(1-p) \sin({\pi p}/2)}{\pi} .
\end{align}
In the case $p=2$ the result is $(\kappa|t-t'|)^{p-1}$, which is equivalent to taking the limit $p\to 2$ in Eq.~\eqref{p1to3}.

For $p\ge 3$ we upper bound the spectrum as
\begin{align}
\tilde\Sigma_0(\omega) &\le \left\{
\begin{array}{ll}
\kappa^{p-1}/\gamma^p, & |\omega| < \gamma,\\
\kappa^{p-1}/|\omega|^p, & |\omega| \ge \gamma.
\end{array}
\right.
\end{align}
In the case $p=3$, we then get
\begin{align}
&\expect\Bk{[X(t)-X(t')]^2} 
\nn &\le \frac{(\kappa|t-t'|)^{2}}{\pi}\log\left(\frac 1{\gamma|t-t'|}\right) +O[(\kappa|t-t'|)^2].
\end{align}
For $p>3$,
\begin{align}
&\expect\Bk{[X(t)-X(t')]^2} \le (\kappa|t-t'|)^{2}\frac{p(\gamma/\kappa)^{3-p}}{3\pi(p-3)} \nonumber \\
&\quad + (\kappa|t-t'|)^{p-1} \frac{2\Gamma[2-p]\sin(\pi p/2)}{\pi(p-1)}+O[(\kappa|t-t'|)^3].
\end{align}
For $|t-t'|=T$, in each case we find that the variance varies as a polynomial in $\kappa/{\cal N}$, and is therefore small for large ${\cal N}/\kappa$.

\section{Justification of approximations in Eq.~\eqref{longapprox}}
\label{corrbou}

Using the definition of $K_n$, we find that
\begin{align}
Z_n &= K_n-K_{n-1} 
= \frac 1{2\pi} [\check X_n - \check X_{n-1}\nn
& \quad -X(nT)+X((n-1)T) -\xi_n+\xi_{n-1}].
\end{align}
Now $Z_n$ is an integer, and 
\begin{equation}
\frac 1{2\pi} |\check X_n - \check X_{n-1}| \le 1/2.
\end{equation}
This means that $[-X(nT)+X((n-1)T) -\xi_n+\xi_{n-1}]/(2\pi)$ takes values within $1/2$ of $Z_n$, which in turn implies that
\begin{align}
\label{eq:rou}
&Z_n = \nn
&{\rm round} \left(\frac 1{2\pi} [-X(nT)+X((n-1)T) -\xi_n+\xi_{n-1}] \right).
\end{align}
Here the rounding is taken to use the round half up convention.
Now define
\begin{align}
C_n &:= [-X(nT)+X((n-1)T)]/(2\pi), \\
D_n &:= [-\xi_n+\xi_{n-1}]/(2\pi),
\end{align}
so
\begin{equation}
Z_n = {\rm round} \left(C_n+D_n \right).
\end{equation}
The important thing to note is that, for $|n-m|>1$, $D_n$ is independent of $D_m$ because these only depend on the error in independent measurements.
Moreover, $D_n$ and $D_m$ are independent of $C_n$ and $C_m$.
However, $C_n$ and $C_m$ can be correlated due to correlations in the phase variation.

Using this notation and expanding $\expect\Bk{ Z_n Z_m }$ in terms of the probability distribution gives
\begin{widetext}
\begin{align}
\label{posnegsum}
\expect\Bk{ Z_n Z_m } &= \sum_{z_n} \sum_{z_m} \Pr(Z_n=z_n,Z_m=z_m) z_n z_m \nn
&= 2\sum_{z_n>0} \sum_{z_m>0} \Pr(Z_n=z_n,Z_m=z_m) z_n z_m + 2\sum_{z_n>0} \sum_{z_m<0} \Pr(Z_n=z_n,Z_m=z_m) z_n z_m \nn
&= 2\sum_{z_n>0} \sum_{z_m>0} \Pr(Z_n\ge z_n,Z_m\ge z_m) - 2\sum_{z_n>0} \sum_{z_m<0} \Pr(Z_n\ge z_n,Z_m\le z_m) \nn
&= 2\sum_{z_n>0} \sum_{z_m>0} \left[\Pr\left(C_n+D_n\ge z_n-1/2,C_m+D_m\ge z_m-1/2\right)  \right.
\nn & \quad \left. - \Pr\left(C_n+D_n\ge z_n-1/2,C_m+D_m\le -z_m+1/2\right) \right].
\end{align}
\end{widetext}

We use this expression to bound $\expect\Bk{ Z_n^2 }$, justifying the approximation used in the last line of Eq.~\eqref{longapprox}.
This also bounds $\expect\Bk{ Z_nZ_m }$, because $\expect\Bk{ Z_n^2 }\ge |\expect\Bk{ Z_nZ_m }|$.
For $n=m$ then $C_n=C_m$, and $D_n=D_m$, and we need only consider $\Pr(C_m+D_m\ge z_m-1/2)$.
For $n=m$, $\Pr(C_n+D_n\ge z_n-1/2,C_m+D_m\le -z_m+1/2)=0$ since the two conditions are incompatible.
If $z_m\ge 2$, then the probability $\Pr(D_m\ge z_m-3/4)$ is zero, so
\begin{equation}
\Pr(C_m+D_m\ge z_m-1/2) \le \Pr(C_m\ge z_m-7/4).
\end{equation}
This is exponentially small in ${\cal N}/\kappa$ \emph{and} $z_m$.
Because this term decays exponentially with $z_m$, the sum over $z_m$ is exponentially small in ${\cal N}$.
As a result, we have that
\begin{equation}
\expect\Bk{ Z_m^2 } \lapprox 2 \Pr(C_m+D_m\ge 1/2) = p_{\rm err},
\end{equation}
where $p_{\rm err}$ is the probability of a phase-wrap error.
Similarly we have $\expect\Bk{ Z_m Z_{m+1} } \lapprox p_{\rm err}$.
This justifies the approximation in the last line of Eq.~\eqref{longapprox}.
Note that
\begin{equation}
\Pr(C_m+D_m\ge 1/2) \le \Pr(C_m\ge 1/4) + \Pr(D_m\ge 1/4).
\end{equation}
The probability $\Pr(C_m\ge 1/4)$ is exponentially small in ${\cal N}$, and can be ignored in comparison to $\Pr(D_m\ge 1/4)$.
This is why $p_{\rm err}$ is approximately equal to the probability of $|\xi_n-\xi_{n-1}|$ exceeding $\pi/2$.

Next we wish to show that the sum omitted in the second-last line of Eq.~\eqref{longapprox} has size exponentially small in ${\cal N}$.
It is relatively straightforward to show that the individual terms in that sum are exponentially small.
The difficulty is in showing that the sum is also exponentially small, since it is over an infinite number of terms.
For $|n-m|>1$, we wish to evaluate the difference of probabilities
\begin{widetext}
\begin{align}
&\Pr(C_n+D_n\ge z_n-1/2,C_m+D_m\ge z_m-1/2)- \Pr(C_n+D_n\ge z_n-1/2,C_m+D_m\le -z_m+1/2)\nn
&= \int dd_n \int dd_m \left[\Pr(C_n\ge z_n-1/2-d_n,C_m\ge z_m-1/2-d_m)\right. \nn
& \quad \left. - \Pr(C_n\ge z_n-1/2-d_n,C_m\le -z_m+1/2+d_m)\right]\Pr(D_n=d_n)\Pr(D_m=d_m) .
\end{align}
In the last line we have used the symmetry of the probability distribution $\Pr(D_m=d_m)$ about zero.
We are interested in the case where the variance in $C_n$ (equal to the variance of $C_m$) is small.
It is small in comparison to $z_n-1/2$, and therefore we can perform an expansion in $1/(z_n-1/2)$, and similarly for $z_m$.
We are also interested in the case where the covariance between $C_n$ and $C_m$ is small, so we also perform an expansion in the covariance about zero.

Let us denote $\sigma^2:=\expect\Bk{ C_n^2}$ and $V_{n-m}:=\expect\Bk{C_n C_m}$.
Then we obtain the expression
\begin{align}
&\Pr(C_n\ge z_n-1/2-d_n,C_m\ge z_m-1/2-d_m) - \Pr(C_n\ge z_n-1/2-d_n,C_m\le -z_m+1/2-d_m) \nn
&\approx \frac{V_{n-m} (z_n-1/2-d_n)(z_m-1/2-d_m)}{\pi\sigma^2(z_n-1/2)(z_m-1/2)}\exp\left[-\frac{(z_n-1/2-d_n)^2+(z_m-1/2-d_m)^2}{2\sigma^2}\right] .
\end{align}

Expanding in a series for $d_n$ and $d_m$ about zero, as these will also be small as compared to $z_n$ and $z_m$, we get
\begin{align}
&\Pr(C_n\ge z_n-1/2-d_n,C_m\ge z_m-1/2-d_m) - \Pr(C_n\ge z_n-1/2-d_n,C_m\le -z_m+1/2-d_m) \nn
&\approx \frac{V_{n-m}}{\pi\sigma^2}\exp\left[-\frac{(z_n-1/2)^2+(z_m-1/2)^2}{2\sigma^2}\right]
\left\{ 1 + \frac{[\sigma^2-(z_m-1/2)^2]d_m}{(z_m-1/2)\sigma^2}+\frac{[(z_m-1/2)^2-3\sigma^2]d_m^2}{2\sigma^4} \right. \nn
&\quad  \left.+\frac{[\sigma^2-(z_n-1/2)^2]d_n}{(z_n-1/2)\sigma^2}
+\frac{[(z_n-1/2)^2-\sigma^2][(z_m-1/2)^2-\sigma^2]d_nd_m}{(z_n-1/2)(z_m-1/2)\sigma^4} + \frac{[(z_n-1/2)^2-3\sigma^2]d_n^2}{2\sigma^4}  \right\} ,
\end{align}
where we have omitted terms higher than second-order in $d_n$ and $d_m$
Also omitting terms of first-order since they will average to zero, as well as $d_nd_m$ since that will average to zero, this simplifies to
\begin{align}
&\Pr(C_n\ge z_n-1/2-d_n,C_m\ge z_m-1/2-d_m) - \Pr(C_n\ge z_n-1/2-d_n,C_m\le -z_m+1/2-d_m) \nn
&\approx \frac{V_{n-m}}{\pi\sigma^2}\exp\left[-\frac{(z_n-1/2)^2+(z_m-1/2)^2}{2\sigma^2}\right]
\left\{ 1 +\frac{[(z_m-1/2)^2-3\sigma^2]d_m^2}{2\sigma^4} +\frac{[(z_n-1/2)^2-3\sigma^2]d_n^2}{2\sigma^4} 
 \right\}.
\end{align}
This expression decays exponentially with $z_n$ and $z_m$, so we can omit terms with $z_n>1$ or $z_m>1$, and obtain
\begin{equation}
\label{expdec}
\expect\Bk{ Z_n Z_m } \approx \frac{V_{n-m}}{\pi\sigma^2}\exp\left[-\frac{1}{4\sigma^2}\right]
\left\{ 1 +\frac{[1/4-3\sigma^2]\expect\Bk{ D_n^2}}{\sigma^4}
 \right\} .
\end{equation}

The crucial feature of this expression is that it varies \emph{linearly} in $V$, and decays \emph{exponentially} with ${\cal N}$ (because $\sigma^2$ decreases polynomially in ${\cal N}$.
Evaluating $V$, we obtain
\begin{align}
V_{n-m} &= \expect\Bk{ C_n C_m } 
= \expect\Bk{ \{X((n-1)T)-X(nT)\}\{X((m-1)T)-X(mT)\}} \nn
&= 2\Sigma_0((n-m)T)-\Sigma_0((n-m+1)T)-\Sigma_0((n-m-1)T) \nn
&= \frac 1{\pi} \int d\omega \, \tilde\Sigma_0(\omega) \cos[\omega(n-m)T]\left[ 1-\cos\omega T \right].
\end{align}

Taking $m=0$ for simplicity, and integrating by parts, we obtain
\begin{align}
V_{n} 
&= -\frac 1{\pi} \int_0^{\infty} d\omega \, \tilde\Sigma'_0(\omega) \left\{\frac 1{nT}\sin(\omega nT)-
\frac 1{2(n-1)T}\sin[\omega (n-1)T]-\frac 1{2(n+1)T}\sin[\omega (n-1)T]\right\},
\end{align}
\end{widetext}
where we have used the fact that $\tilde\Sigma_0(\omega)$ is bounded at $\omega=0$ and approaches zero for $\omega\to\infty$.
Using the fact that $\tilde\Sigma'_0(\omega)\le 0$,
\begin{align}
&|V_{n}| \nn
&\le -\frac 1{\pi} \int_0^{\infty} d\omega \, \tilde\Sigma'_0(\omega) \left\{\frac 1{nT}+
\frac 1{2(n-1)T}+\frac 1{2(n+1)T}\right\} \nn
&= \frac 1{\pi} \tilde\Sigma_0(0)\left\{\frac 1{nT}+\frac 1{2(n-1)T}+\frac 1{2(n+1)T}\right\}
\end{align}

As $\tilde\Sigma_0(0)$ is bounded (equal to $\kappa^{p-1}/\gamma^p$), $V_{n-m}$ scales as $1/|n-m|$.
Hence, $\expect\Bk{ Z_n Z_m }$ scales as a factor exponentially small in ${\cal N}$ times $1/|n-m|$.

Next we consider the scaling for $a_m$.
Using the first two terms of the asymptotic series for the digamma function,
\begin{equation}
\dig(x)=\ln(x)-\frac 1{2x}+O(x^{-2}),
\end{equation}
we have
\begin{align}
&\dig(x+1/2)-\dig(x)\nn
&= \ln(x+1/2)-\ln(x)-\frac 1{2x+1}+\frac 1{2x}+O(x^{-2}) \nn
&=\frac 1{2x}+O(x^{-2}).
\end{align}
As a result, $a_m$ has the scaling
\begin{equation}
a_m = -(-1)^m \sin\left(\frac{\pi t}{T}\right)
\frac 1{|m|} + O(m^{-2}).
\end{equation}

Now we have sufficient results to bound the component of the sum
\begin{equation}
S= \sum_{n,m:|n-m|>1} \expect\Bk{ Z_n Z_m } a_m a_n
\end{equation}
that was omitted in Eq.~\eqref{longapprox}.
Using the above results the sum can be bounded as
\begin{equation}
S\lapprox f({\cal N})\sum_{n,m:|n-m|>1,n\ne 0,m\ne 0} \frac 1{|n-m|\times |n|\times |m|},
\end{equation}
where $f$ is an exponentially decreasing function.
Here we have only included the leading-order terms in the asymptotic expansion, because the higher-order terms will result in higher powers in the denominator, which give smaller results.
Splitting the sum into $m<n$ and $m>n$, the bound can be rewritten as
\begin{equation}
S \lapprox 4f({\cal N})\sum_{m>n>1} \frac 1{|n-m| n m} = 4f({\cal N})\sum_{n,r>0} \frac 1{n(n+r)r}.
\end{equation}
Using the inequality $x^2+y^2\ge 2xy$ for
$x = \sqrt{n}$ and $y = \sqrt{r}$  gives
$ n+ r \ge 2\sqrt{nr}$.
Hence, substituting $n+r$ with $2\sqrt{nr}$ gives
\begin{align}
S &\lapprox 4f({\cal N})\sum_{n,r>0} \frac 1{n (2\sqrt{nr}) r} \nn
&= 2 f({\cal N}) \sum_{n,r>0} \frac{1}{n^{3/2} r^{3/2}} \nn
&= 2 f({\cal N}) [\zeta(3/2)]^2 \nn
&\le 14 f({\cal N}),
\end{align}
where $\zeta(z)$ is the Riemann zeta function.
This means that the sum is exponentially small in ${\cal N}$, which is why it can be omitted in Eq.~\eqref{longapprox}.

\end{document}